\documentclass[12pt]{iopart}

\usepackage{graphicx}
\usepackage{blindtext, stfloats, overpic, subfig}

\usepackage{float}
\usepackage{tabularx,booktabs}
\newcolumntype{C}{>{\centering\arraybackslash}X} 
\usepackage{lipsum}

\pdfoutput = 1

\begin{document}
	
	\title[]{New Approach for Designing cVEP BCI Stimuli Based on Superposition of Edge Responses}
	
	\author{Muhammad Nabi Yasinzai\textsuperscript{1} and Yusuf Ziya Ider\textsuperscript{2}}
	\address{\textsuperscript{1,2}Department of Electrical and Electronics Engineering, Bilkent University, Ankara Turkey}
	\ead{\textsuperscript{1}muhammad.nabi.yasinzai@gmail.com, \textsuperscript{2}ider@ee.bilkent.edu.tr}

	\vspace{10pt}
	\begin{indented}
		\item[]February 2020
	\end{indented}
	
	\begin{abstract}
		The purpose of this study is to develop a new methodology for designing stimulus sequences for cVEP BCI based on experimental studies regarding the behavior and the properties of the actual EEG responses of the visual system to coded visual stimuli, such that training time is reduced and possible number of targets is increased. EEG from 8 occipital sites is recorded with 2000 samples/sec per channel, in response to visual stimuli presented on a computer monitor with 60Hz refresh rate. Onset and offset EEG responses to long visual stimulus pulses are obtained through 160-trial signal averaging. These edge responses are used to predict the EEG responses to arbitrary stimulus sequences using the superposition principle. A BCI speller which utilizes the target templates generated by this principle is also implemented and tested. It is found that, certain short stimulus patterns can be accurately predicted by the superposition principle. BCI sequences that are constructed by combinations of these optimal patterns yield higher accuracy (95.9\%) and ITR (57.2 bpm) compared to when the superposition principle is applied to conventional m-sequences and randomly generated sequences. Training time for the BCI application involves only the acquisition of the edge responses and is less than 4 minutes, and a huge number of sequences is possible. This is the first study in which cVEP BCI sequences are designed based on constraints obtained by observing the actual brain responses to several stimulus patterns.
	\end{abstract}
	
	\vspace{2pc}
	
	%
	%
	
	\section{Introduction}

	BCI based on visual evoked potentials (VEPs), which were first proposed in 1984 as a new communication channel \cite{sutter1984visual} \cite{nagel2018modelling}, are the most common type of EEG based BCIs. In 1992, code modulated VEP (cVEP) was introduced and 64-bit long binary stimulus patterns (known as m-sequences) are employed for the task of target classification \cite{sutter1992brain}. A base m-sequence is assigned to one target (base target), and the sequences for the remaining targets are circularly shifted versions of the base sequence. A signal-averaged EEG response “template” is recorded for the base target, and the templates for the rest of the targets are generated by circular shifting of the recorded EEG template. In BCI application EEG is recorded while the subject gazes at a certain target. The recorded EEG is correlated with the templates of all targets, and the target that yields maximum correlation is accepted. In m-sequence based cVEP, only a limited number of shifts is possible based on the length of the stimulus pattern. For instance, a 63-bit long m-sequence is widely used for cVEP based BCI spellers (Visual Keyboard). In these spellers, the m-sequence is circularly shifted by 2-bits for each target and therefore, they can support a maximum of 32 targets. To increase the number of targets, it is necessary to either increase the length of the sequence or decrease the number of bits for shifting, which will decrease the speed or accuracy of the BCI speller, respectively. Another viable solution is to combine multiple targets into a single group and use a specific m-sequence for each group \cite{wei2018novel}. However, this method is limited since there is a small number of acceptable m-sequences for a specific sequence length \cite{wei2018novel} \cite{nagel2018modelling} and training is need for each of the m-sequences.
	
	Training time is also an important factor to consider when assessing BCIs. M-sequence has the advantage of less training time because different templates are obtained by shifting. Even though shifted m-sequences may themselves be mutually orthogonal, their EEG responses may not be as orthogonal \cite{bacsaklar2019effects}. To circumvent this problem, one may consider different codes for each target which are not shifted versions of each other. This choice however increases the training time. Therefore, it is necessary to be able to obtain templates for different targets from a small set of training data. One notable study in these lines is the work done by Nagel et. al. \cite{nagel2018modelling}, who have investigated the use of arbitrary code sequences for cVEP BCI. A moving average (MA) model is proposed in his study to predict brain responses to arbitrary stimuli patterns. Some training data are used to estimate the coefficients of the MA model and using this model all target templates are predicted. The accuracy and Information Transfer Rate (ITR) values of the BCI system which uses these templates are reported to be high. 
	
	Nagel et. al. work is based to a large extent on the linearity of the system. However, it is known from modeling studies  \cite{robinson2003nonuniform} that the visual system is nonlinear. The 13 parameter Robinson’s model has 4 sigmoid shaped nonlinear blocks in addition to gain and linear filter blocks \cite{robinson2003nonuniform}. Therefore the use of linear models have to be handled with caution. The fact that the linear approach employed by Nagel et. al. has nevertheless some significant success calls for investigating the nature of the visual system with respect to the linear vs nonlinear perspective. In particular, it is important to investigate under what circumstances the system obeys the superposition and shift-invariance properties. If such properties are indeed satisfied, at least to some degree, then one may predict templates for many stimuli patterns using superposition of responses to simple inputs, which has the advantage of shorter training time.
	
	BCI research is mostly application-specific, where the aim is to come up with a faster and more reliable means of communication interface. Majority of the BCI systems are designed without considering the actual nature of the brain responses. Even some of the new studies which have focused on modeling the brain responses use random stimuli patterns along with some generic models to approximate EEG responses to them \cite{nagel2018modelling} \cite{nagel2019world}. Therefore, the concept of designing a BCI system based on the nature and properties of the brain responses to visual stimuli is still untouched.
	
	In this paper, we investigate the possibility of using the superposition principle for predicting the brain responses to different stimuli patterns. Initially, the EEG responses for simple stimulus patterns are acquired, which are then superimposed to generate the EEG responses to more complex stimuli patterns. Since we have found that the brain responds to the edges of the code sequence, as shown in the results section, we have paid attention to obtaining the edge responses correctly using averaging. In fact, we call this phase as the training phase. We then performed several studies, using more general stimulus sequences, to find the similarity between the generated (simulated) EEG signals and the measured (observed) EEG signals. It was observed that the generated and observed EEG had fairly high correlation, provided that some of the constraints on the structure of the stimuli sequences are met. Later in this paper, a BCI system based on the superposition property is proposed, implemented, and tested.
	
	\section{Materials and Methods}

	\subsection{Participants}
	Seven participants (7 male, mean age of 29.7 years, and standard deviation of 5.5 years) participated in the experiment after providing a written consent form approved by the ethical committee of Bilkent University. They had normal or corrected-to-normal vision, had no previous history of epilepsy, and they were informed about the objective of this study.

	\subsection{Visual Speller Design Methods}
\begin{figure}[!t]
		\centering
		\includegraphics[width=5in]{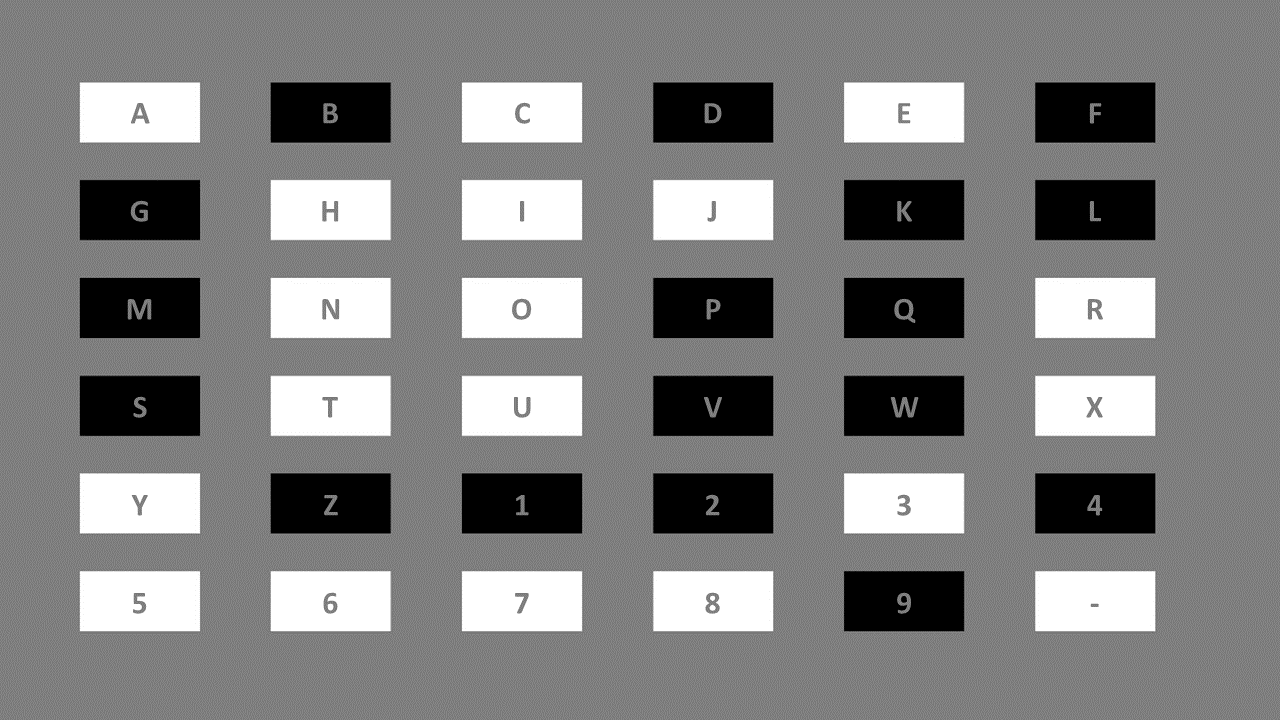}
		\caption{Visual Speller Screen for the BCI experiments.}
		\label{fig_sim}
	\end{figure}

	The visual speller screen is designed in Matlab environment using Psychtoolbox \cite{brainard1997psychophysics} and is presented on a 25-inch LED (Dell Alienware AW2518HF) at 60 Hz refresh rate with a resolution of 1920 x 1080 pixels. Subjects were seated in front of the monitor screen at a distance of about 80 cm. The visual speller consists of 36 targets, arranged as a 6 x 6 matrix on the screen (Fig. \ref{fig_sim}). Each target has a rectangular shape of 180 x 90 pixels (5.18cm x 2.6cm) with the respective letter/number written at its center. The targets include letters from A to Z, numbers from 1 to 9 and a “-” symbol.  The targets are toggled to black or white when the corresponding stimulus code is 0 and 1, respectively. Each bit of the stimulus sequence is presented for 16.667ms (display time of a single frame).
	
	Ubuntu 18.04 with low latency kernel was used in the “stimulus computer” for having accurate stimulus timings. Close attention was given to the Psychtoolbox’s missed-frames counter to make sure that no frames were dropped during an experiment. A PIN photodiode circuit, explained in detail in our previous study \cite{tuncel2018period}, was used for measurement of the actual refresh rate of the monitor, and it was found to be 59.94 Hz. For the purpose of synchronization, a marker pulse is transmitted from the stimulus computer to the EEG amplifier after the update of each monitor frame to keep track of the flashing time for each bit. We also measured the exact delay between the frame update and the marker pulse. A negligible delay of  280 $\mu s$ was found \cite{memon2019hybrid}. 
	
	\subsection{Data Acquisition}
	Brain Products, V-Amp-16 EEG amplifier is used for recording EEG, with the standard 10-20 EEG cap which has 32 electrode locations (Brain Products, Gilching, Germany). V-Amp-16 contains a total of 16 EEG amplifier channels, however, in our experiments EEG is recorded from 8 channels of the amplifier with electrodes placed at positions ‘O1, Oz, O2, Pz, P3, P4, P7 and P8’. The reference electrode is placed on FCz position, and the ground electrode is located on the forehead above nasion. Active/wet electrodes are used and the electrode impedances are measured using ImpBox (Brain Products, Gilching, Germany) and are kept below 10k$\Omega$. The sampling rate of the EEG amplifier is set to 2000 sps per channel. BCI2000 \cite{schalk2004bci2000}  with FieldTrip \cite{oostenveld2011fieldtrip} are used for recording EEG responses along with the synchronization markers, to another computer (the "recording computer") using MATLAB session in real time.
	
	\subsection{Data Preprocessing}
	A 4-40 Hz bandpass filter is applied to all of the EEG signals along with a 50 Hz notch filter to remove any of the remaining 50 Hz interference. EEG responses of the stimuli sequences are averaged over the trials with the help of the synchronization markers. Furthermore, the signals are then spatially averaged using the coefficients determined by canonical correlation analysis (CCA) \cite{spuler2013spatial}. CCA helps to increase the signal to noise ratio and reduces the 8-channel data into a single signal.

	\subsection{Stimulus Sequences }
\begin{figure}[!t]
		\centering
 		\includegraphics[width=5in]{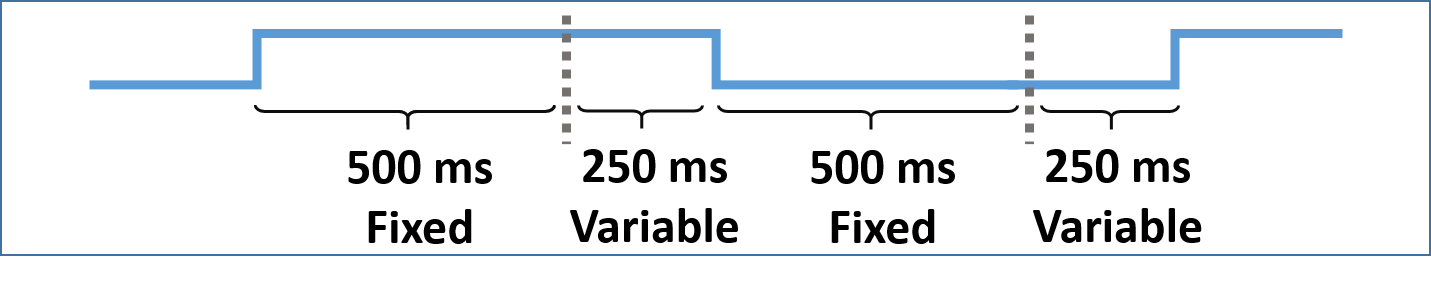}
		\caption{Wide pulse stimulus with randomized duration of the high and low regions.}
		\label{aquireEdges}
	\end{figure}

	The experiments are divided into three types. In the first type, EEG responses to wide stimuli pulses are acquired. In the second type of experiments, EEG responses to 9 different stimulus sequences with simple (short) patterns are obtained. In the final experiment type, various long BCI sequences are studied, and a BCI speller is realized and tested. The stimulus pattern for the first experiment type is given in Fig. \ref{aquireEdges}. Each time this stimulus pattern is repeated, the lengths of the high and low regions are randomly chosen between 500 ms and 750 ms to remove any artifacts that may arise due to the otherwise periodic nature of the EEG. The stimulus pattern in Fig. \ref{aquireEdges} is assigned to target A and repeated 160 times for averaging. The other targets are assigned randomly generated sequences and the subject is asked to gaze at target A.  
	
\begin{figure}[!t]
		\centering
		\includegraphics[width=5in]{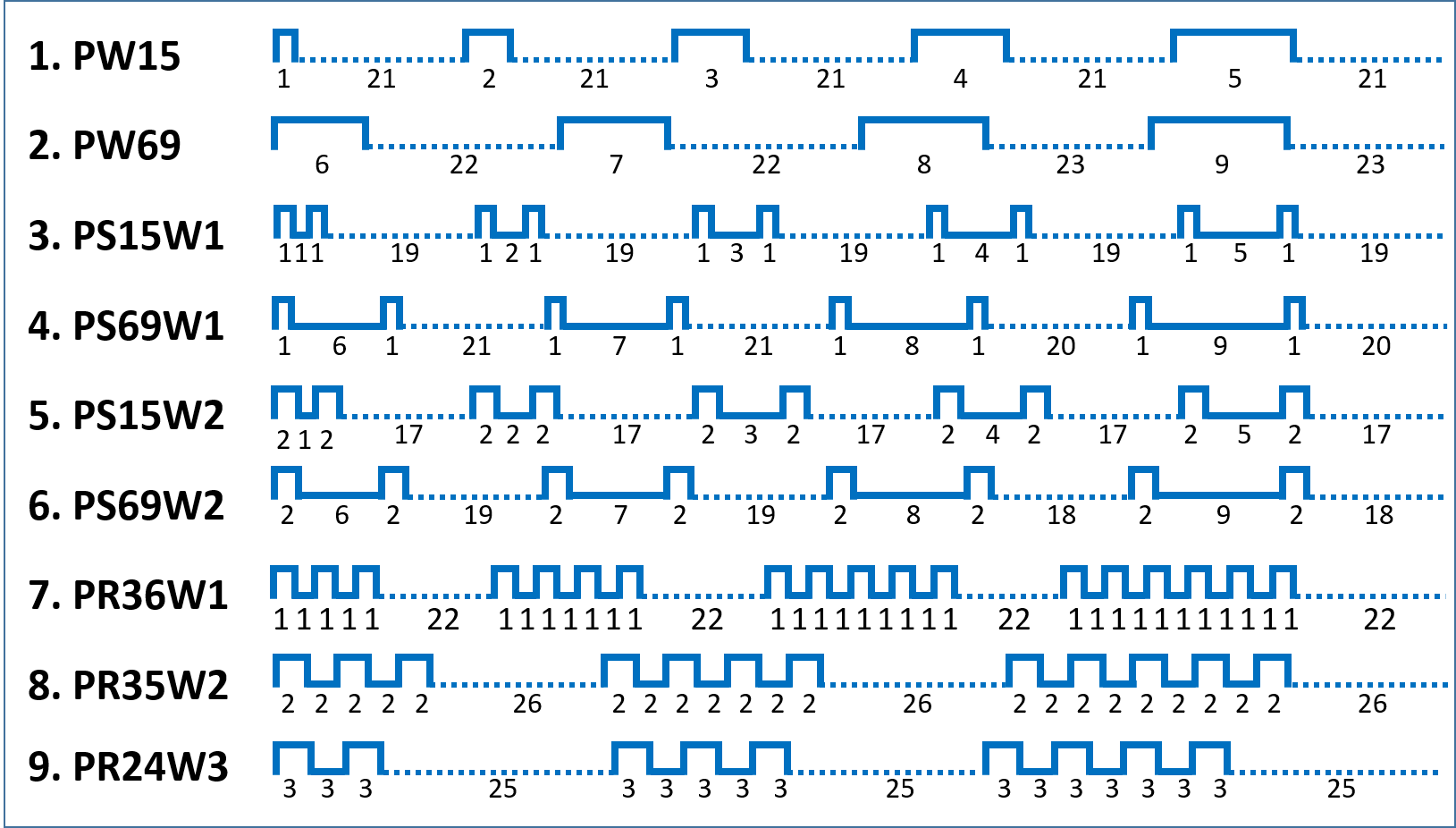}
		\caption{ 9 different stimuli sequences designed to investigate the nature of EEG responses to simple visual stimulus patterns.
		}
		\label{stimulusSequences}
	\end{figure}

	For the second type of experiments, the stimulus sequences shown in Fig. \ref{stimulusSequences} are used to carry out the following studies; Pulse Width (PW) study, Pulse Separation (PS) study, and Pulse Repetition (PR) study. In the PW study, stimulus sequences PW15 and PW69 are used, which cover pulse widths of 1-5 and 6-9 bits, respectively. In Fig. \ref{stimulusSequences}, the widths of each part of the stimulus sequences are written below in units of bits, and they add up to 120 bits. The PS study is divided into 2 parts. In the first part, the separation between two 1-bit pulses is changed between 1 to 5 bits (sequence PS15W1) and between 6 to 9 bits (sequence PS69W1). In the second part, the separation between two 2-bit pulses is changed between 1 to 5 bits (sequence PS15W2) and between 6 to 9 bits (sequence PS69W2). Finally, in the PR study, different repetitions of 1-bit, 2-bit, and 3-bit wide pulses are studied. In this study, the distance between the pulses is equal to the width of the pulses (pulses are periodically repeated with 50\% duty cycle). The stimulus sequences PR36W1, PR35W2 and PR24W3 cover 3-6  repetitions of 1-bit wide pulses, 3-5  repetitions of 2-bit wide pulses, and 2-4  repetitions of 3-bit wide pulses, respectively. It should be noted that PR36W1 and PR35W2 sequences do not include the case of 2 repetitions of 1-bit and 2-bit pulses because they are already covered in the PS study. 
	
	The solid lines in Fig. \ref{stimulusSequences} are the patterns under consideration, whereas the dotted lines are the separations between these patterns. Separations of 17-26 bits (283 ms - 433 ms) are introduced, so that the EEG responses to the individual patterns do not overlap. For each of the 9 stimulus sequences, EEG is recorded for 30 trials for signal averaging. These 9 stimuli sequences are assigned to target A, and the subject is asked to gaze at target A while the remaining targets are assigned randomly generated sequences. 
	
\begin{figure}[!t]
		\centering
		\includegraphics[trim={2.5cm 1.2cm 1cm 0cm},clip,width=5in]{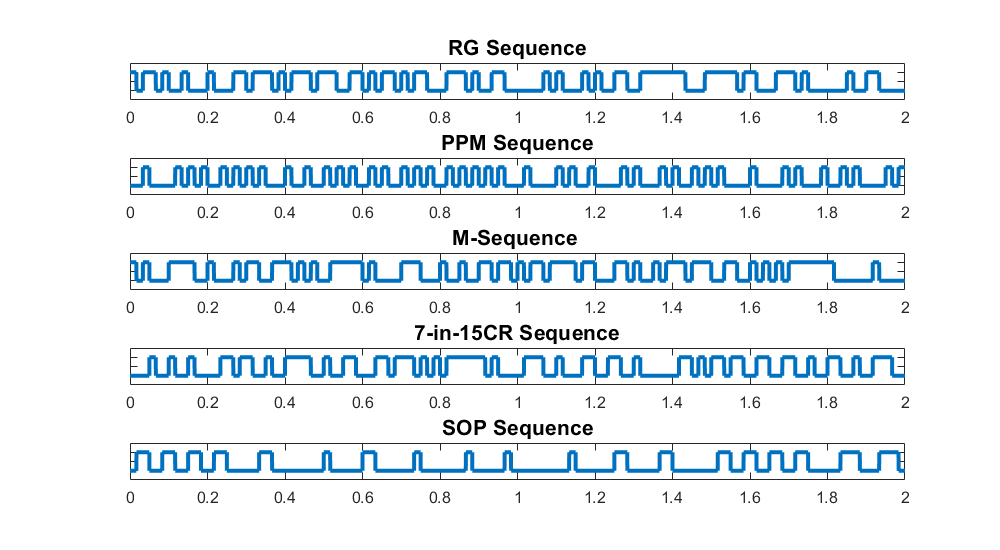}
		\caption{Different Types of 120-bit stimulus sequences }
		\label{allStimulusFigure}
	\end{figure}

	In the last type of experiments, we studied 5 different 120-bit long sequences which are potentially to be used in BCI experiments (Fig. \ref{allStimulusFigure}). These sequences are named as Randomly Generated (RG) Sequence, Pulse Position Modulated (PPM) Sequence, m-sequence, 7-in-15 Change Random (7-in-15CR) Sequence and Superposition Optimized Pulse (SOP) Sequence. RG Sequence is obtained by assigning 1 or 0 to every next bit position with 50\% probability. The PPM sequence contains 1-bit pulses which are separated randomly by a distance of 1 to 4 bits. The 120-bit m-sequence is actually a truncated version of a 127-bit m-sequence, discussed in our previous paper \cite{bacsaklar2019effects}. The 7-in-15CR sequence contains 7 changes for every 15 bits of the sequence. This code was proposed by Nagel et. al. \cite{nagel2018modelling} and was reported to have a better performance than randomly generated codes. Finally, the SOP sequence is proposed by us in the result section. To generate such a code we first selected 16 different small sequence patterns. They include 1-bit pulse followed by 5-10 zero bits, 2-bit pulse followed by 5-10 zero bits, 3 and 4 repetitions of 2-bit pulses followed by 5 zero bits and finally, 3 and 4 repetitions of 3-bit pulses followed by 5 zero bits. Hence, there are 16 different patterns to choose from, and an SOP sequence is generated by randomly picking up these patterns, concatenating them, and truncating the final pattern to 120 bits. 
	
	\subsection{BCI Training and Testing}
	The training phase of our BCI speller involves obtaining the average EEG responses to wide pulses shown in Fig. \ref{aquireEdges}. These responses are then used to predict the template EEG responses for all targets using the superposition based procedure, as explained in the results section. During the test phase, the subjects are asked to spell 35 targets in sequence from “A-Z” and “1-9”, respectively. The EEG responses are recorded for 2 trials and are averaged. During testing, the acquired EEG is correlated with the generated EEGs (templates) of all targets, and the target which has the maximum correlation is selected.

	\begin{equation*}
		ITR  = \frac{60}{T} \times (log_2N + P log_2P + (1-P)log_2 \frac{1-P}{N-1}) \label{eq:itrEq}
	\end{equation*}
	
	BCI accuracy is defined as the percentage of the number of targets correctly classified. ITR is another standard performance metric and is defined as in equation \ref{eq:itrEq}, where, P is the classifier accuracy, N is the total number of targets and T is the time required to classify a single target. Two trials of the 120-bit stimulus (4 seconds) plus the gaze shifting time (1 second) make T equal to 5 seconds.

	\section{Results}
	
	\subsection{EEG responses to wide pulses}
\begin{figure}[!t]
		\centering
		\includegraphics[trim={3cm 2cm 1.8cm 1.5cm},clip,width=5in]{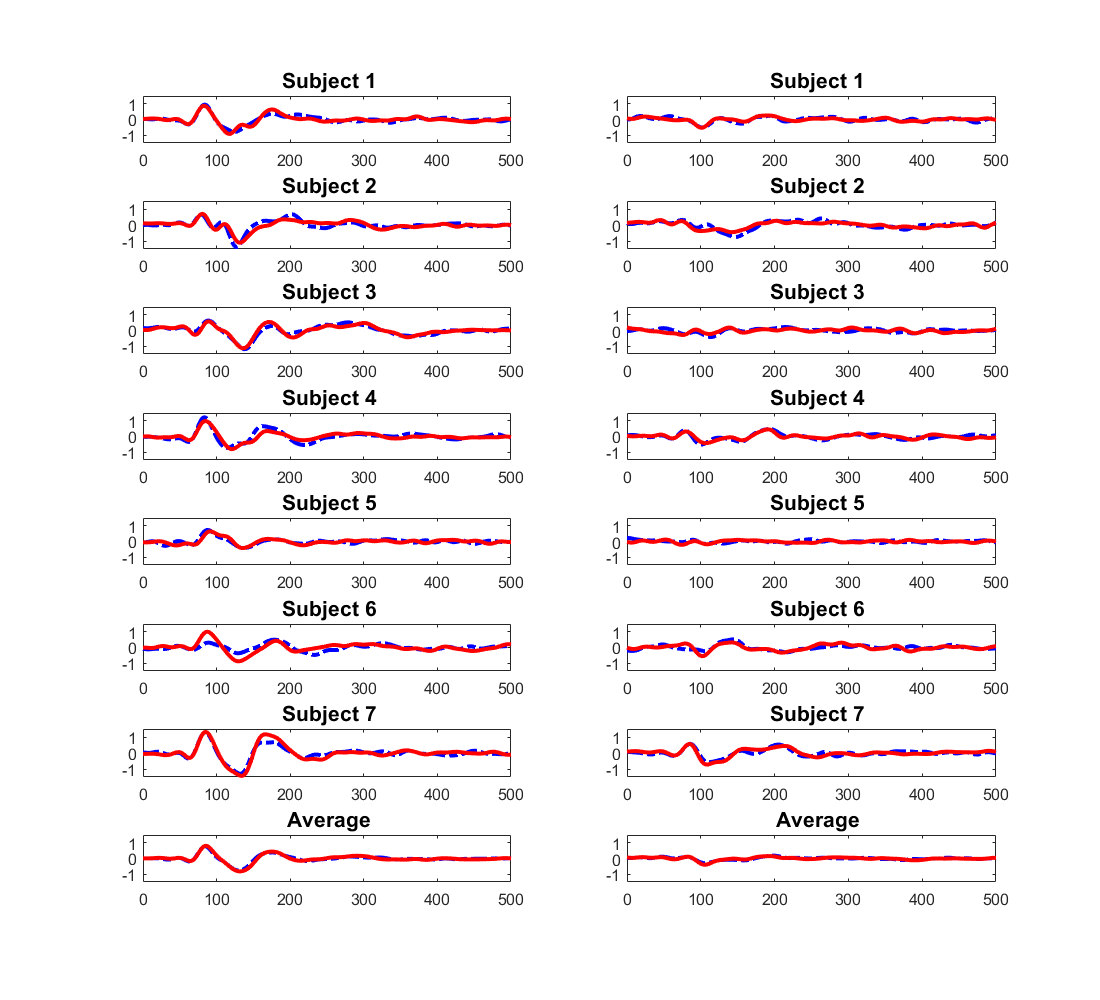}
		\caption{Onset and Offset responses are on the Left and Right, respectively. Dotted-blue lines are the edge responses obtained 2 week after the first set of responses shown in red.}
		\label{posNegEdgeAndRepetability}
	\end{figure}

	Fig. \ref{posNegEdgeAndRepetability} shows that the 160-repetition averaged EEG responses of each of the 7 subjects start 50 ms after either the positive edge (onset) or the negative edge (offset) of the stimulus and wane within 350 ms. It can be observed that each subject has his own distinct onset and offset responses, but they do not deviate significantly from the averaged responses of all 7 subjects. Fig. \ref{posNegEdgeAndRepetability} also shows the edge responses acquired 2 weeks later, indicating that onset and offset responses are repeatable. The onset responses acquired from the two acquisitions have correlations of 0.90, 0.82, 0.93, 0.86, 0.81, 0.61, and 0.96 for subjects 1-7, respectively. For the offset responses, the corresponding correlations are 0.79, 0.80, 0.44, 0.82, 0.04, 0.67 and 0.87. The relatively lower correlation values for the offset responses are due to lower signal to noise ratio because of the low amplitudes of the offset responses. The average RMS value of the onset responses for acquisitions 1 and 2 are 0.303 and 0.289, respectively, whereas the corresponding average RMS values of offset responses are 0.149 and 0.160.

	\subsection{Using superposition for predicting cVEP responses}
	If the system is linear and if the system responds to the edges only, then it should be possible to predict the EEG response to a general stimulus pattern by superposition, that is, by shifting the onset and offset response to the positions of the corresponding edges of the stimulus sequence and adding them. In the following, we investigate to what extent the superposition principle is valid. Correlation between the predicted and the actual responses are then used to evaluate the performance of the superposition-based procedure first for the different simple pulse patterns provided in Fig. \ref{allStimulusFigure}, and then also for long BCI sequences.
	
	\subsubsection{Prediction Performance for different short Pulse Widths:}
\begin{figure}[!t]
		\centering
		\includegraphics[trim={4.2cm 1.1cm 3cm 0.5cm},clip,width=5in]{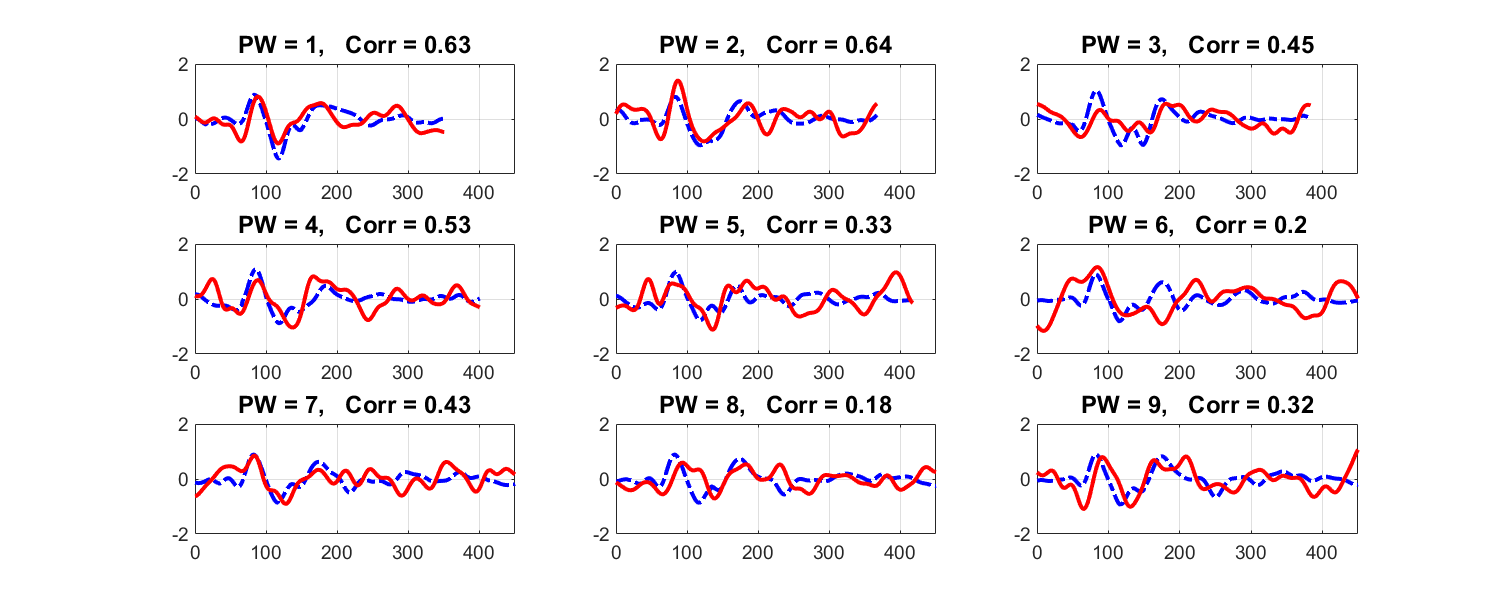}
		\caption{Subject 1, Acquired (Red-Solid) and Generated (Dotted-Blue) EEG responses for 1-9 bit wide pulses.  }
		\label{pw_responses}
	\end{figure}

	In this PW study, the EEG responses for different pulse widths are acquired using the stimulus patterns PW15 and PW69. The observed and generated (predicted) EEG responses for different pulse widths are given in the Fig. \ref{pw_responses} for Subject 1, and it is observed that as the pulse width is increased from 1 to 9 bits the correlation falls from 0.63 to 0.32. The average correlations, over all 7 subjects, between the generated and the observed EEGs for different stimulus pulse widths are illustrated in Fig. \ref{avgCI_figure}a. For PWs of 1 and 2 bits the correlations are around 0.7 and are lower for the other PWs. Repeated measures ANOVA test shows that there is significant difference among the correlations obtained for different PWs (p $<$  0.001). When paired t-test was applied to pairwise compare different PWs (p $<$ 0.05), we observed that the correlations for 1-bit and 2-bit wide pulses are similar, but they are significantly different from most of the other PWs. Therefore, we arrive at the understanding that using 1-bit and 2-bit wide pulses in a stimulus sequence will give better accuracy between the predicted and the acquired EEG responses. (The individual correlations of all 7 subjects and the detailed results of the statistical test are provided in the Supplementary Information. The same holds for the other prediction performance studies reported in the following).
	
	\subsubsection{Prediction Performance for different Pulse Separations:}
\begin{figure}[!t]
		\centering
		\includegraphics[trim={4.2cm 1.1cm 3cm 0.5cm},clip,width=5in]{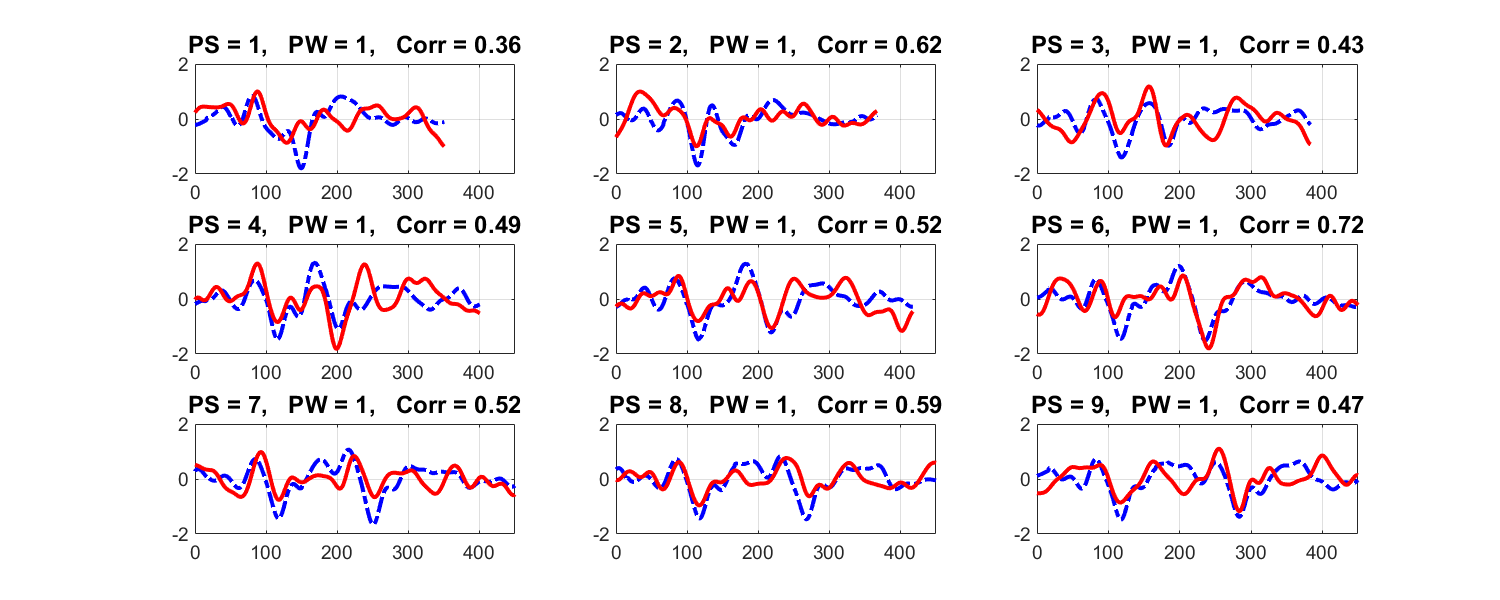}
		\caption{Subject 1, Acquired (Red-Solid) and Generated (Dotted-Blue) EEG responses for stimulus patterns of 1-9 bit separation between 1-bit wide pulses.}
		\label{psw1_responses}
	\end{figure}

	In this PS study, the effect of time separation between adjacent pulses is studied. The pulses of interest are either 1-bit wide or 2-bit wide pulses. First, the separation distance between two adjacent 1-bit pulses is changed from 1 to 9 bits and the observed signals for subject 1 are given in Fig. \ref{psw1_responses}. It is found that as the separation is increased from 1 bit to 9 bits the correlation between the measured and predicted (generated) EEG responses increases from 0.36 (for 1-bit separation) to a maximum of 0.72 (for 6-bit separation ) and then gradually falls to 0.47 (for 9-bit separation). The average correlations, over all 7 subjects, between the generated and the observed EEGs for different separations of 1-bit wide pulses are illustrated in Fig. \ref{avgCI_figure}b. On the average, correlation is around 0.4 for separations of 1-3 bits and for higher separations the correlations are around 0.6. Therefore, in general for 1-bit wide pulses if the separation between neighboring pulses is 4 bits or more, then the EEG can be predicted with high accuracy, which is logical because then the overlap between the responses of the individual pulses becomes less. Repeated measures ANOVA test shows that there is a significant difference among the correlations obtained for different pulse separations (p = 0.013). Furthermore, pairwise comparisons using paired t-test shows that the correlations for 4-9 bit separations are statistically not different, whereas correlations for 1-3 bit separations are significantly different from column 6. Therefore, the choice of 4 to 9 bit separations between 1-bit wide pulses seems reasonable.
	
\begin{figure}[!t]
		\centering
		\includegraphics[trim={4.2cm 1.1cm 3cm 0.5cm},clip,width=5in]{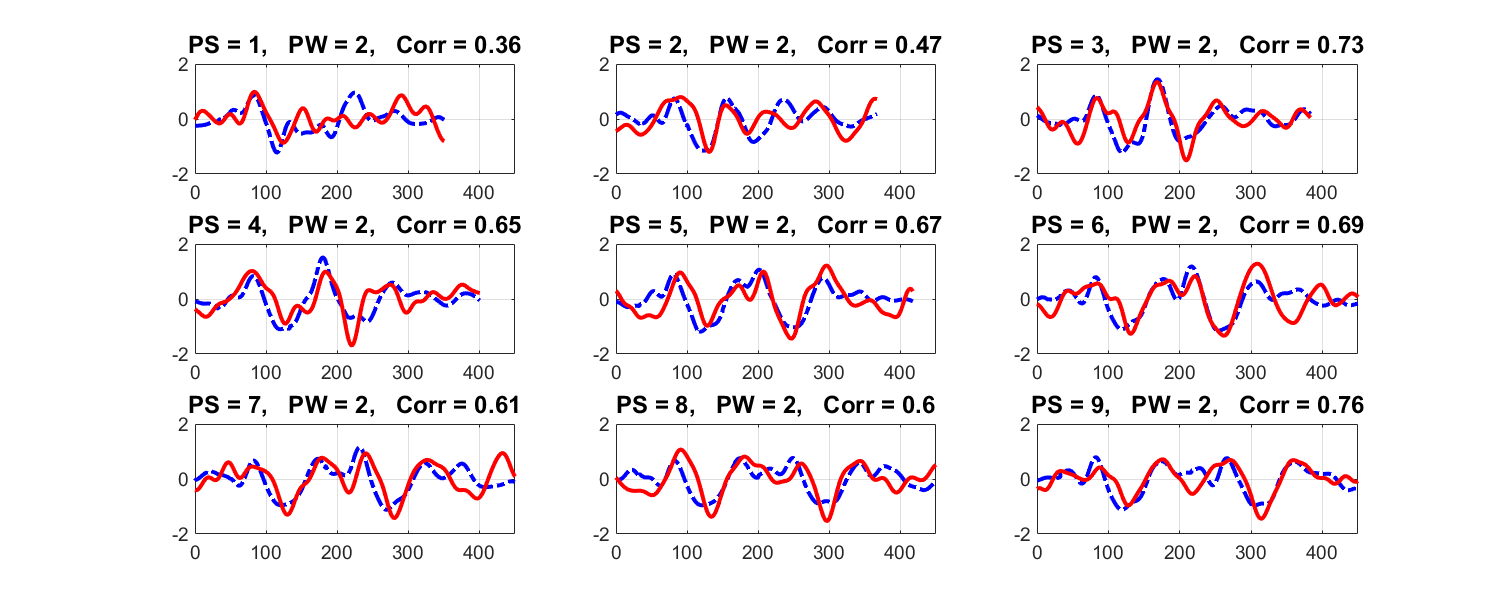}
		\caption{Subject 1, Acquired (Red-Solid) and Generated (Dotted-Blue) EEG responses for stimulus patterns of 1-9 bit separation between 2-bit wide pulses. }
		\label{psw2_responses}
	\end{figure}

	Similarly for 2-bit wide pulses, the same pattern is observed. The generated and observed EEG plots for different pulse separations for Subject 1 are provided in Fig. \ref{psw2_responses}. The average correlations, over all 7 subjects, between the generated and the observed EEGs for different separations of 2-bit wide pulses are illustrated in Fig. \ref{avgCI_figure}c. The average correlation increases from 0.32 to 0.65 as separation is increased from 1 to 3 bits. From 3-bit up to 9-bit separations small variation in the correlations are observed. Hence, it can be inferred that if the separation between the 2-bit wide pulses is greater than or equal to 3 bits, the responses of the individual pulses will have less overlap and the generated sequence response will have a high correlation with the observed EEG response. Repeated measures ANOVA test shows that the null hypothesis of having equal means for different separations is to be rejected (p = 0.001). Furthermore, pairwise comparisons using paired t-test shows that the mean correlations for 3-9 bit separations are similar to each other and the mean correlations for 1-bit and 2-bit separations are significantly different from 3, 4 and 9 bit separations. Therefore in general choice of 3-bit to 9-bit separation between 2-bit pulses seems reasonable.

	\subsubsection{Prediction Performance for different Pulse Repetitions:}
\begin{figure}[!t]
		\centering
		\includegraphics[trim={4.2cm 1.1cm 3cm 0.5cm},clip,width=5in]{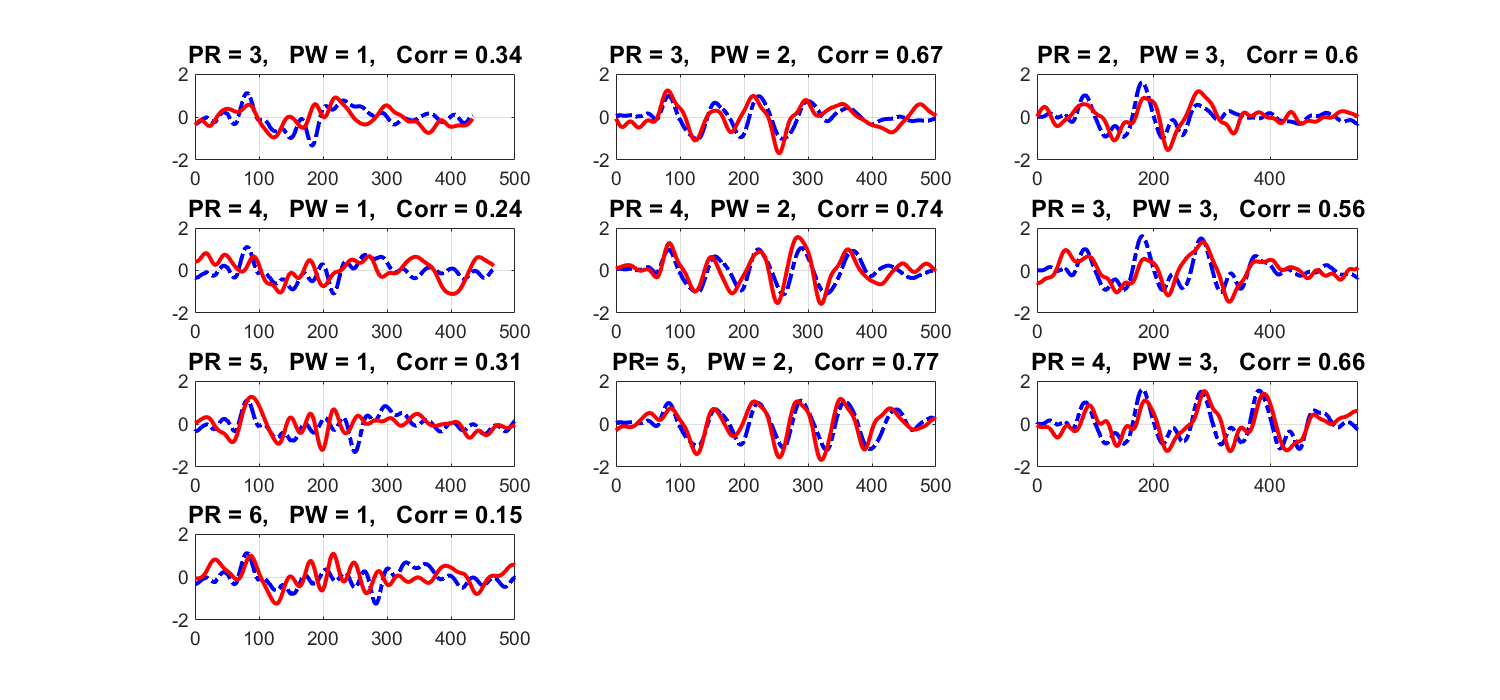}
		\caption{Subject 1, Acquired (Red-Solid) and Generated (Dotted-Blue) EEG for pulse repetitions of 1-3 bit wide pulses. }
		\label{prRes}
	\end{figure}

	In this PR study, the generated and measured EEG are compared for different number of repetitions of 1-bit, 2-bit, and 3-bit wide pulses. The stimulus sequences used for this study are PR36W1, PR35W2 and PR24W3 for 1, 2 and 3-bit pulses, respectively, as shown in Fig. \ref{stimulusSequences}.  The results of this study for subject 1 is shown in Fig. \ref{prRes}. For 1-bit pulse, the correlation for 3 repetitions is 0.34, which then drops to 0.15 as the number of repetitions is increased to 6. This is because of the destructive interaction of the individual edge responses, which can also be observed in Fig. \ref{prRes} by the low amplitude of the acquired EEG. Therefore, it appears that repetitions of 1-bit pulses are not to be preferred. For 2 repetitions of 2-bit pulses the correlation is less than 0.5 (as seen From Fig. \ref{psw2_responses}). However for 3, 4 and 5 repetitions, the correlations are between 0.67 and 0.77. For 3-bit wide pulses, the correlation for 2, 3 and 4 repetitions is between 0.56 and 0.66. Therefore, 3-5 and 2-4 repetitions of 2-bit and 3-bit wide pulses give better accuracy results, respectively.

\begin{figure}[!t]
		\centering
		\includegraphics[trim={0cm 0cm 1cm 0cm},clip,width=5in]{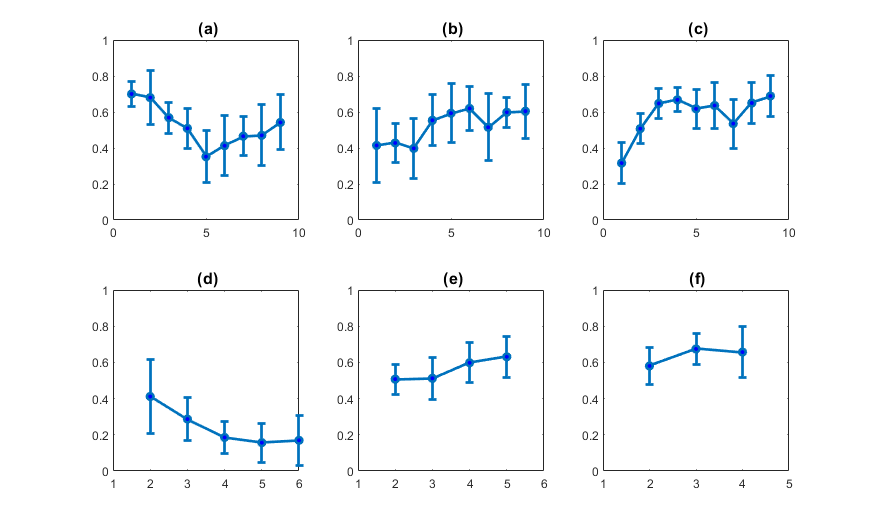}
		\caption{Average Correlation values with 95\% confidence interval. (a) For 1-9 Pulse Widths, (b) For 1-9 Separations between 1-bit wide pulses, (c) For 1-9 Separations between 2-bit wide pulses, (d) For 2-6 Repetitions of 1-bit wide pulse, (e) For 2-5 Repetitions of 2-bit wide pulse, (f) For 2-5 Repetitions of 3-bit wide pulse,  }
		\label{avgCI_figure}
	\end{figure}

	The average correlations over all 7 subjects and their confidence intervals for different repetitions of 1-bit, 2-bit and 3-bit wide pulses are illustrated in Fig. \ref{avgCI_figure}d, \ref{avgCI_figure}e and \ref{avgCI_figure}f, respectively. Overall, the results for all subjects are in parallel to the results we have explained for subject 1. Repeated measures ANOVA indicates that significant difference exists between different repetitions of 1-bit pulses (p = 0.03) but pairwise paired t-tests do not show any significant difference between different repetitions (at 5\% significance level) and also the average correlation values for all repetitions are very low (less than 0.411). Therefore, repetition of 1-bit pulses is not preferred. On the other hand, the correlation for 2-bit wide pulses shows much higher values and this phenomenon in observed in almost all of the subjects. Repeated measures ANOVA indicates that correlations for different repetition are significantly different (p $<$ 0.01). Pairwise paired t-test results indicate that although for all repetition cases the average correlations are higher than 0.507, the correlation for 4 and 5 repetitions are significantly higher than 2 and 3 repetition cases. Hence for 2-bit wide pulses, 4 and 5 repetitions are to be preferred. For the repetition of 3-bit pulses, it is found from repeated measures ANOVA that all repetition cases are statistically the same (p = 0.16) and the average correlations are higher than 0.58. Pairwise comparisons also do not show any significant difference between the correlations of different repetition cases of 3-bit pulses. Hence, 3-bit wide pulses will give good correlation between the predicted and the measured EEGs, for 2-4 repetitions. 
	
	\subsubsection{Prediction Performance for long Stimulus Sequences:}
\begin{figure}[!t]
		\centering
		\includegraphics[trim={2cm 1cm 1.5cm 0.5cm},clip,width=5in]{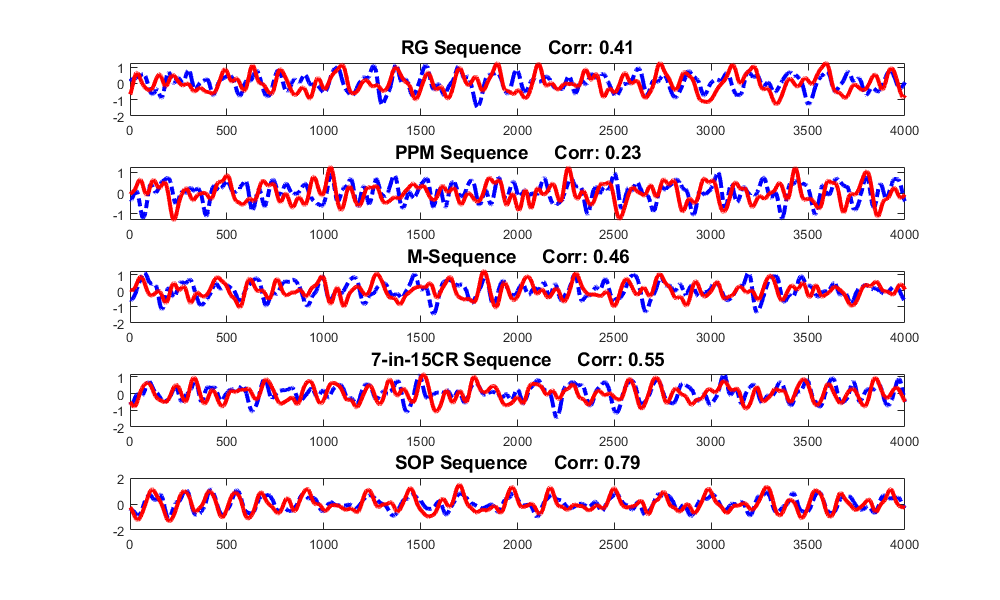}
		\caption{Generated and Observed EEG responses for a) RG Sequence, b) PPM Sequence, c) M-Sequences, d) 7-in-15CR Sequences, and e) SOP. }
		\label{Sub1_5_Gen_Obs_responses}
	\end{figure}

	In this study we tested, for subject 1, the 5 different stimulus sequences which are provided in Fig. \ref{allStimulusFigure} of section II.E. All of these sequences are 120 bits long, they are repeated for 60 trials to get a good averaged signal, and the results are shown in Fig. \ref{Sub1_5_Gen_Obs_responses}. The correlation between the generated and actual EEG responses for the RG sequence, PPM sequence, m-sequence, 7-in-15CR sequence, and the SOP sequence are 0.41, 0.23, 0.46, 0.55, and 0.79, respectively. The EEG response to the PPM sequence is difficult to predict because the separations between 1-bit pulses are very small. The correlation values for the m-sequence and the RG sequence are also low, probably because of the same reason. The 7-in-15CR sequence performed better than the RG, PPM and m-sequences. Finally, our proposed SOP sequence performed the best, and a high correlation was recorded between the generated and observed EEGs.
	
	In the following, a BCI application is proposed, which uses target code sequences which are designed by taking the above results into consideration. In fact, the "atomic pulse waveforms" which are used to design the proposed SOP stimulus sequences, which are explained in section II.E (Stimulus Sequences) are decided upon as a result of the aforementioned observations. Acquiring BCI results for all of the 5 sequence types shown in Fig. \ref{allStimulusFigure} would be very demanding for the subjects, and therefore we decided to compare the least performing PPM sequence, the moderately performing 7-in-15CR sequence, and the best performing SOP sequence.

	\subsection{BCI Results}
	Table \ref{bci_table} provides the accuracy and ITR values of the BCI application using the three stimulus types for each subject. For all of the subjects, the PPM sequences performed the worst with an overall accuracy and ITR of 6.94\% and 1.7 bits/min, respectively. The performance of the 7-in-15CR sequences is comparatively better than the PPM sequence with an average accuracy and ITR of 27.75\% and 10.53 bits/min, respectively. Finally, as expected, the SOP sequences performed the best with an accuracy of 95.9\% and ITR of 57.19\%, respectively. It is worth repeating that the PPM sequences performed the worst because in these sequence only 1-bit wide pulses are used and the separation between the pulses are randomly chosen between 1 and 4 bits. As we have shown in section “Prediction Performance for different Pulse Separations”, the pulse separation should be greater than 3 bits for better prediction accuracy. Also in the 7-in-15CR sequences there are many instances when pulse separations are short, and this may explain why these sequences also do not perform well. In general we may conclude, although from the comparison results of just 3 sequence types, that sequences which do not obey the rules that we have identified in the previous sections, perform poorly.
	
	\begin{table}[!t]
		\centering
		\caption{BCI Application Results}
		\label{bci_table}
		\begin{tabular}{|c|c|c|c|c|c|c|}
			\hline
			\textbf{} & \multicolumn{2}{c|}{\textbf{PPM}} & \multicolumn{2}{c|}{\textbf{7-in15CR}} & \multicolumn{2}{c|}{\textbf{SOP}} \\ \hline
			\textbf{\begin{tabular}[c]{@{}c@{}}Sub \\ No\end{tabular}} & \textbf{\begin{tabular}[c]{@{}c@{}}Acc\\ (\% )\end{tabular}} & \textbf{\begin{tabular}[c]{@{}c@{}}ITR  \\ (bits/min)\end{tabular}} & \textbf{\begin{tabular}[c]{@{}c@{}}Acc\\ (\% )\end{tabular}} & \textbf{\begin{tabular}[c]{@{}c@{}}ITR \\ (bits/min)\end{tabular}} & \textbf{\begin{tabular}[c]{@{}c@{}}Acc\\ (\% )\end{tabular}} & \textbf{\begin{tabular}[c]{@{}c@{}}ITR\\ (bits/min)\end{tabular}} \\ \hline
			\textbf{01} & 37.14 & 11.93 & 94.29 & 54.73 & 100 & 62.04 \\ \hline
			\textbf{02} & 0 & 0 & 14.29 & 2.18 & 100 & 62.04 \\ \hline
			\textbf{03} & 2.86 & 0 & 22.86 & 5.25 & 100 & 62.04 \\ \hline
			\textbf{04} & 0 & 0 & 20.0 & 4.13 & 91.4 & 51.67 \\ \hline
			\textbf{05} & 5.71 & 0.21 & 17.14 & 3.10 & 88.54 & 48.82 \\ \hline
			\textbf{06} & 2.86 & 0 & 20.0 & 4.13 & 100 & 62.04 \\ \hline
			\textbf{07} & 0 & 0 & 5.71 & 0.21 & 91.4 & 51.67 \\ \hline
			\textbf{Avg} & \textbf{6.94} & \textbf{1.7} & \textbf{27.75} & \textbf{10.53} & \textbf{95.9} & \textbf{57.19} \\ \hline
		\end{tabular}
	\end{table}

	\section{Discussion}
	The human brain is a highly nonlinear and complex system, and yet, most of the cVEP based applications are designed without considering the characteristics of the brain responses. In m-sequence based cVEP BCI applications, the assumption of shift-invariance seems to hold. Similarly, the BCI application based on an MA model [13] also performed quite well.  Therefore, it can be inferred from these studies that modeling or just making some assumptions on the characteristics of the system may be useful in designing BCI applications. In our study, we have gone one step further to understand the system better and have investigated how well the brain complies with the superposition property. This is the first study to investigate the nature of brain responses to different visual stimulus patterns to verify that the brain responses follow superposition under certain constraints on the stimulus patterns. The observations are then used for designing optimum sequences for BCI.
	
	The constraints which we are suggesting for BCI stimulus sequences decrease the number of possible targets. However, it is still a huge set, and a large number of targets can be introduced. Furthermore, we have performed experiments on only a few simple stimulus patterns, and the patterns that performed well among them are used in designing the long stimulus sequences for our BCI application. Further studies can be carried out in order to identify additional predictable simple patterns, and they can be added to the set of acceptable simple patterns to increase the number of targets. 
	
	The positive and negative edges of the stimulus sequence are the main factors influencing the EEG response. These positive and negative responses have an initial delay of 50 ms and diminish 350 ms after the edge. These responses appear to be similar to the step response of a low order, such as  2nd or 3rd order, linear system. Hence, further studies can be carried out to obtain a model of the system by estimating the parameters of such a system from the edge responses.
	
	The edge responses of the seven subjects in our study have a common pattern, it is worth investigating if a universal edge response can be used for predicting responses to BCI target sequences. Using a universal edge response would have the advantage of eliminating the training stage for acquiring the edge responses of an individual subject. However, at this stage, until further studies are undertaken, we suggest that a training session should be carried out for each individual to acquire his/her edge responses.
	
	The results that we have obtained not only serve for better design of BCI experiments but also shine light on the workings of the visual system regarding its linearity and shift-invariance properties. We hope that our results will give insight to researchers who deal with the fundamental aspects of the visual system in addition to investigators who undertake application-oriented research such as BCIs.

	\section{Conclusion}
	In this study, responses of the visual system to several code patterns are first studied in order to come up with rules (constraints) for constructing beneficial sequences for BCI applications. It is first found that the onset and offset responses of the brain to visual stimulus pulses are delayed by about 50 ms and wane within 350 ms. These edge responses are then used to predict EEG responses to any stimulus sequences by using superposition. It is found that 1-bit and 2-bit wide pulses with 4-9 bit and 3-9 bit separations respectively, and 2-bit and 3-bit pulses with 3-4 and 2-4 repetitions respectively, can be predicted with good accuracy. These simple patterns were randomly combined in 120-bit stimulus sequences to be used in a BCI speller application. It is confirmed that with such target stimulus sequences, the BCI application will give better classification results with 95.9\% accuracy. Furthermore, the BCI application proposed in our study has short training time because the training is carried out only to acquire the edge responses, and the proposed methodology allows for a large number of possible targets. The results of our study indicate that although the visual system is known to be nonlinear; nevertheless, based on some simple constraints on the structure of the stimulus sequences, a linear operation like superposition can be used in a BCI application.

	\section*{References}
	\bibliographystyle{iopart-num}
	\bibliography{References}

\providecommand{\newblock}{}
\begin{thebibliography}{10}
\expandafter\ifx\csname url\endcsname\relax
  \def\url#1{{\tt #1}}\fi
\expandafter\ifx\csname urlprefix\endcsname\relax\def\urlprefix{URL }\fi
\providecommand{\eprint}[2][]{\url{#2}}

\bibitem{sutter1984visual}
Sutter E~E 1984 The visual evoked response as a communication channel {\em
  Proceedings of the IEEE Symposium on Biosensors\/} pp 95--100

\bibitem{nagel2018modelling}
Nagel S and Sp{\"u}ler M 2018 {\em PloS one\/} {\bf 13} e0206107

\bibitem{sutter1992brain}
Sutter E~E 1992 {\em Journal of Microcomputer Applications\/} {\bf 15} 31--45

\bibitem{wei2018novel}
Wei Q, Liu Y, Gao X, Wang Y, Yang C, Lu Z and Gong H 2018 {\em IEEE
  Transactions on Neural Systems and Rehabilitation Engineering\/} {\bf 26}
  1178--1187

\bibitem{bacsaklar2019effects}
Ba{\c{s}}aklar T, Tuncel Y and Ider Y~Z 2019 {\em Biomedical Physics \&
  Engineering Express\/} {\bf 5} 035023

\bibitem{robinson2003nonuniform}
Robinson P, Whitehouse R and Rennie C 2003 {\em Physical Review E\/} {\bf 68}
  021922

\bibitem{nagel2019world}
Nagel S and Sp{\"u}ler M 2019 {\em bioRxiv\/}  546986

\bibitem{brainard1997psychophysics}
Brainard D~H 1997 {\em Spatial vision\/} {\bf 10} 433--436

\bibitem{tuncel2018period}
Tuncel Y, Ba{\c{s}}aklar T and Ider Y~Z 2018 {\em Biomedical Physics \&
  Engineering Express\/} {\bf 4} 025024

\bibitem{memon2019hybrid}
Memon S~A 2019 {\em Hybrid and model based approaches for new BCI spellers\/}
  Ph.D. thesis Bilkent University

\bibitem{schalk2004bci2000}
Schalk G, McFarland D~J, Hinterberger T, Birbaumer N and Wolpaw J~R 2004 {\em
  IEEE Transactions on biomedical engineering\/} {\bf 51} 1034--1043

\bibitem{oostenveld2011fieldtrip}
Oostenveld R, Fries P, Maris E and Schoffelen J~M 2011 {\em Computational
  intelligence and neuroscience\/} {\bf 2011} 1

\bibitem{spuler2013spatial}
Sp{\"u}ler M, Walter A, Rosenstiel W and Bogdan M 2013 {\em IEEE Transactions
  on Neural Systems and Rehabilitation Engineering\/} {\bf 22} 1097--1103

\end{thebibliography}

	\pagebreak

	\setcounter{table}{0}
	\renewcommand{\thetable}{S\arabic{table}}
	
	\setcounter{figure}{0}
	\renewcommand{\thefigure}{S\arabic{figure}}
	
	\setcounter{figure}{0}
	\renewcommand{\thefigure}{S\arabic{figure}}
	
	\setcounter{section}{0}
	\renewcommand{\thesection}{S\arabic{section}}

	\title[]{SUPPLEMENTARY MATERIAL}

	\section{Illustration of the idea of using superposition for predicting cVEP responses}
\begin{figure}[H]
		\centering
		\includegraphics[width=5in]{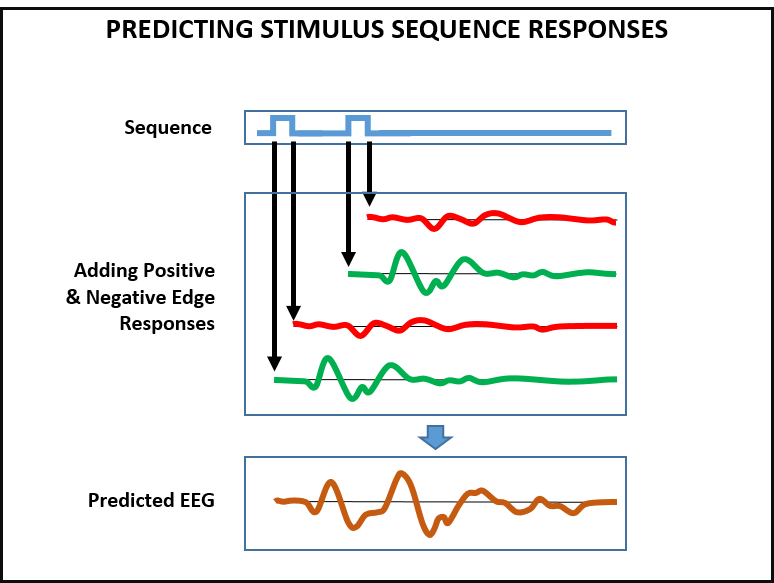}
		\caption{The blue waveform is the stimulus pattern, and it consists of 2 pulses only. The green and red signals are the onset and offset responses that are shifted according to the location of the corresponding edges in the stimulus sequence. These shifted edge responses are added to get the predicted EEG (orange signal).}
		\label{superposModel}
	\end{figure}

	The process of predicting EEG response to a stimulus pattern employs shifting of the edge responses to the corresponding location of these edges in the pattern and then adding these shifted edge responses. This superposition-based procedure is illustrated in Figure \ref{superposModel}.

	\section{Prediction Performance for different short Pulse Widths}
	In this PW study, the EEG responses for different pulse widths are acquired using the stimulus patterns PW15 and PW69. The correlation values for each of the 7 subjects for is given in Table \ref{pwStudy}. For all subjects except subject 5, the maximum correlation is obtained either for 1-bit or 2-bit pulse widths. In general, it can be noted that as the pulse width is increased from 1-bit the correlation between the generated EEG and observed EEG decreases and reaches a minimum correlation of 0.35 for 5-bit wide pulse and then increase slowly for 5-9 bits of pulse width. 
	
	Table \ref{pwStudyANOVA} provides the results of the Repeated Measure ANOVA test for this study. The small p-value (p $<$ 0.001) shows that the correlations obtained for different pulse widths are significantly different.
	
	Table \ref{pwPairWiseResult} provides the results of pairwise paired t-tests between the correlation values of 1-9 bit wide pulses. In this Table, a value of "1" in any cell means that the correlation values are significantly different, and "0" shows that the correlation values are similar to each other at 5$\%$ significance level (The same is applied for the pairwise paired t-test for the remaining studies). The correlation values obtained for 1-bit wide pulses have a significant difference from the correlation values of 3-8 bit pulse widths, but the correlations for 1-bit and 2-bit wide pulses are similar to each other. Table \ref{pwPairWisePval} provides the actual p-values for the pairwise paired t-tests.
	
	\section{Prediction Performance for different separations between 1-bit pulses}
	In the first part of this PS study, EEG responses for 1-9 bit separations between two 1-bit wide pulses are studied. The correlation for all of the 7 subjects are provided in Table \ref{ps1Table}. The average correlation over all subjects is around 0.3 for 1-3 bit separations between the pulses, whereas the correlation is around 0.6 for 4-5 bit separation between the pulses. 
	
	Table \ref{ps1ANOVA} provides the results of the Repeated Measure ANOVA test. This test shows that the null hypothesis of having equal means for the columns of Table \ref{ps1Table} is to be rejected (p = 0.013). 
	
	Table \ref{ps1PairWisePResult} gives the pairwise comparisons results between the correlations obtained for 1-9 bit separations between 1-bit wide pulses using paired t-test. This test clearly shows that the correlations obtained for 4-9 bit separations are similar. Table \ref{ps1PairWisePval} provides the actual p-values for the pairwise paired t-tests.
	
	\section{Prediction Performance for different separations between 2-bit pulses}
	In the second part of this PS study, EEG responses for 1-9 bit separations between two 2-bit wide pulses are studied.  The correlations for all of the 7 subjects are provided in Table \ref{ps2Table}. In general the average correlation over all subjects is less than 0.51 for pulse separations of 1-2 bits, and the correlation is above 0.54 for 3-9 bits of separations.
	
	Table \ref{ps2ANOVA} provides the results of the repeated measures ANOVA test and it shows that the null hypothesis of having equal means for the columns of Table \ref{ps2Table} is to be rejected (p $<$ 0.001). 
	
	Table  \ref{ps2PairWisePResult} provides the pairwise comparisons using paired t-tests. The correlations for 3-9 bit separations are similar to each other, but 1-2 bit separations have significant difference from 3, 4 and 9 bit separations between 2-bit wide pulses. Furthermore, Table \ref{ps2PairWisePval} provides the actual p-values of the pairwise paired t-tests.

	\section{Prediction Performance For 2-6 Repetitions of 1-bit wide pulse}
	Table \ref{prW1Table} provides the correlation values between the generated and acquired EEG responses for 2-6 repetitions of 1-bit wide pulse. In general the correlation is low for all repetitions of 1-bit wide pulse (below 0.41). 
	
	Table \ref{prW1ANOVA} gives the results for the repeated measures ANOVA test. This test shows that there is a significant difference between the different repetitions of 1-bit wide pulse (p = 0.03). 
	
	The results of the pairwise paired t-tests are given in Table \ref{prW1tTest}. These tests do not show any significant difference between the correlations obtained for 1-6 repetitions of 1-bit wide pulse. The detailed p-values for the pairwise paired t-tests are given in Table \ref{prW1Pvalues}.

	\section{Prediction Performance For 2-5 Repetitions of 2-bit wide pulse}
	Table \ref{prW2Table} provides the correlation values between the generated and acquired EEG responses for 2-5 repetitions of 2-bit wide pulse. The average correlation, over all subjects, is around 0.5 for 1 and 2 repetitions and it is around 0.6 for 3 and 4 repetitions of 2-bit wide pulse.
	
	Table \ref{prW2ANOVA} provides the results of the repeated measures ANOVA test. This test indicates that correlations obtained for 2-5 repetitions of 2 bit wide pulse are significantly different (p $<$ 0.01). 
	
	The results of the pairwise paired t-tests are given in Table \ref{prW2tTest}. The results indicate that the correlations for 4 and 5 repetitions are significantly higher than 2 and 3 repetition cases. The actual p-values for the pairwise paired t-tests are provided in Table \ref{prW2pVal}.

	\section{Prediction Performance For 2-4 Repetitions of 3-bit wide pulse}
	Table \ref{prW3Table} provides the correlation values between the generated and acquired EEG responses for 2-4 repetitions of 3-bit wide pulse. The average correlation, over all subjects, is above 0.58 for 2-4 repetitions of the 3-bit wide pulse.
	
	Table \ref{prW3ANOVA} provides the results for the repeated measures ANOVA test. This test indicates that correlations for different 2-4 repetitions of 3-bit wide pulse are not significantly different (p = 0.158). 
	
	The results for the pairwise paired t-tests are given in Table \ref{prW3tTest}. The results indicate that correlations obtained for 2-4 repetitions of 3-bit wide pulses are statistically similar. The actual p-values for the pairwise paired t-tests are provided in Table \ref{prW3pVal}

	\begin{table*}[!htbp]  
		\caption{Correlation values between the generated and recorded EEG responses of 1-9 bit wide pulses for each of the 7 subjects (PW study).}
		\label{pwStudy}
		\begin{tabularx}{\textwidth} {@{}l*{8}{C}c@{}}
			\toprule
			\textbf{Subject No}  & \textbf{1} & \textbf{2} & \textbf{3} & \textbf{4} & \textbf{5} & \textbf{6} & \textbf{7} & \textbf{8} & \textbf{9}    \\ \midrule
			\textbf{1} & 0.634 & 0.643 & 0.448 & 0.530 & 0.326 & 0.197 & 0.435 & 0.179 & 0.316 \\ 
			\textbf{2} & 0.736 & 0.757 & 0.676 & 0.277 & 0.261 & 0.528 & 0.555 & 0.544 & 0.678 \\ 
			\textbf{3} & 0.641 & 0.836 & 0.517 & 0.573 & 0.467 & 0.566 & 0.430 & 0.560 & 0.563 \\ 
			\textbf{4} & 0.767 & 0.468 & 0.532 & 0.553 & 0.319 & 0.143 & 0.414 & 0.303 & 0.357 \\ 
			\textbf{5} & 0.599 & 0.447 & 0.493 & 0.504 & 0.088 & 0.458 & 0.290 & 0.409 & 0.662 \\ 
			\textbf{6} & 0.793 & 0.795 & 0.647 & 0.463 & 0.427 & 0.392 & 0.468 & 0.588 & 0.477 \\ 
			\textbf{7} & 0.731 & 0.812 & 0.668 & 0.664 & 0.574 & 0.606 & 0.667 & 0.710 & 0.742 \\ \midrule
			\textbf{(Avg $\pm$ std)}   & 0.70$\newline\pm$0.07 & 0.68\newline$\pm$0.16 & 0.57\newline$\pm$0.09 & 0.51\newline$\pm$0.12 & 0.35\newline$\pm$0.16 & 0.41\newline$\pm$0.18 & 0.47\newline$\pm$0.12 & 0.47\newline$\pm$0.18 & 0.54\newline $\pm$0.17   \\ 
			\bottomrule
		\end{tabularx}
	\end{table*}
	\begin{table*}[!htbp] 
		\caption{Correlation values between the generated and recorded EEG responses for 1-9 bit separations between two 1-bit wide pulses, for all 7 subjects (PS study).}
		\label{ps1Table}
		\begin{tabularx}{\textwidth} {@{}l*{8}{C}c@{}}
			\toprule
			\textbf{Subject No}  & \textbf{1} & \textbf{2} & \textbf{3} & \textbf{4} & \textbf{5} & \textbf{6} & \textbf{7} & \textbf{8} & \textbf{9}    \\ \midrule
			\textbf{1} & 0.335 & 0.619 & 0.426 & 0.489 & 0.517 & 0.724 & 0.520 & 0.587 & 0.473 \\
			\textbf{2} & 0.699 & 0.449 & 0.274 & 0.415 & 0.468 & 0.671 & 0.559 & 0.596 & 0.728 \\ 
			\textbf{3} & 0.481 & 0.489 & 0.733 & 0.682 & 0.765 & 0.701 & 0.759 & 0.703 & 0.753 \\ 
			\textbf{4} & 0.091 & 0.473 & 0.222 & 0.577 & 0.313 & 0.374 & 0.507 & 0.511 & 0.561 \\ 
			\textbf{5} & 0.507 & 0.321 & 0.230 & 0.336 & 0.688 & 0.593 & 0.258 & 0.455 & 0.507 \\ 
			\textbf{6} & 0.601 & 0.378 & 0.406 & 0.595 & 0.591 & 0.533 & 0.265 & 0.639 & 0.387 \\ 
			\textbf{7} & 0.166 & 0.266 & 0.495 & 0.783 & 0.812 & 0.744 & 0.739 & 0.697 & 0.821 \\  \midrule
			\textbf{(Avg$\pm$ std)} & 0.41\newline$\pm$0.22 & 0.43\newline$\pm$0.12 & 0.40\newline$\pm$0.18 &	0.55\newline$\pm$0.15 & 0.59\newline$\pm$0.18 & 0.62\newline$\pm$0.13 & 0.52\newline$\pm$0.20 &	0.60\newline$\pm$0.09 & 0.60\newline$\pm$0.16 \\ \bottomrule
		\end{tabularx}
	\end{table*}

	\begin{table*}[!htbp]  
		\caption{Correlation values between the generated and recorded EEG responses for 1-9 bit separations between 2-bit wide pulses, for all 7 subjects (PS study).}
		\label{ps2Table}
		\begin{tabularx}{\textwidth} {@{}l*{8}{C}c@{}}
			\toprule
			\textbf{Subject No}  & \textbf{1} & \textbf{2} & \textbf{3} & \textbf{4} & \textbf{5} & \textbf{6} & \textbf{7} & \textbf{8} & \textbf{9}    \\ \midrule
			\textbf{1} & 0.358 & 0.475 & 0.731 & 0.650 & 0.669 & 0.695 & 0.611 & 0.598 & 0.761 \\
			\textbf{2} & 0.238 & 0.431 & 0.477 & 0.619 & 0.694 & 0.596 & 0.671 & 0.641 & 0.771 \\ 
			\textbf{3} & 0.360 & 0.634 & 0.574 & 0.632 & 0.661 & 0.718 & 0.704 & 0.513 & 0.588 \\ 
			\textbf{4} & 0.409 & 0.605 & 0.693 & 0.623 & 0.403 & 0.345 & 0.361 & 0.725 & 0.656 \\ 
			\textbf{5} & 0.282 & 0.505 & 0.702 & 0.608 & 0.515 & 0.659 & 0.450 & 0.554 & 0.482 \\ 
			\textbf{6} & 0.473 & 0.518 & 0.674 & 0.779 & 0.733 & 0.734 & 0.349 & 0.643 & 0.737 \\ 
			\textbf{7} & 0.093 & 0.385 & 0.682 & 0.762 & 0.648 & 0.704 & 0.602 & 0.877 & 0.827 \\ \midrule
			\textbf{(Avg $\pm$ std)}& 0.32\newline$\pm$0.13 & 0.51\newline$\pm$0.09 & 0.65\newline$\pm$0.09 & 0.67\newline$\pm$0.07 &	0.62\newline$\pm$0.12 & 0.64\newline$\pm$0.14 & 0.54\newline$\pm$0.15 & 0.65\newline$\pm$0.12 &	0.69\newline$\pm$0.12 \\ \bottomrule
			
		\end{tabularx}
	\end{table*}

	\begin{table*} [!htbp]  
		\caption{Correlation values between the generated and recorded EEG responses of 2-6 repetitions of 1-bit wide pulse, for all 7 subjects (PR study). }
		\label{prW1Table}
		\begin{tabularx}{\textwidth} {@{}l*{4}{C}c@{}}
			\toprule
			\textbf{Subject No}  & \textbf{2} & \textbf{3} & \textbf{4} & \textbf{5} & \textbf{6}   \\ \midrule 
			\textbf{1} & 0.335 & 0.342 & 0.238 & 0.308 & 0.150 \\
			\textbf{2} & 0.679 & 0.372 & 0.074 & 0.094 & 0.045 \\ 
			\textbf{3} & 0.481 & 0.404 & 0.328 & 0.011 & 0.090 \\ 
			\textbf{4} & 0.091 & 0.416 & 0.077 & 0.239 & 0.064 \\ 
			\textbf{5} & 0.507 & 0.197 & 0.249 & 0.156 & 0.306 \\ 
			\textbf{6} & 0.601 & 0.177 & 0.116 & 0.029 & 0.093 \\ 
			\textbf{7} & 0.166 & 0.101 & 0.216 & 0.259 & 0.440 \\ \midrule
			Correlation(Avg $\pm$ std) &   0.41$\pm$0.22 & 0.29$\pm$0.13  &  0.19$\pm$0.10 & 0.18$\pm$0.12 & 0.17$\pm$0.15    \\ \bottomrule
		\end{tabularx}
	\end{table*}
	
	\begin{table*}[!htbp]
		\caption{Correlation values between the generated and recorded EEG responses of 2-5 repetitions of 2-bit wide pulse, for all 7 subjects (PR study). }
		\label{prW2Table}
		\begin{tabularx}{\textwidth} {@{}l*{3}{C}c@{}}
			\toprule
			\textbf{Subject No}  & \textbf{2} & \textbf{3} & \textbf{4} & \textbf{5}   \\ \midrule
			\textbf{1} & 0.475 & 0.675 & 0.743 & 0.769 \\ 
			\textbf{2} & 0.431 & 0.540 & 0.516 & 0.551 \\ 
			\textbf{3} & 0.634 & 0.456 & 0.517 & 0.647 \\
			\textbf{4} & 0.605 & 0.644 & 0.733 & 0.699 \\
			\textbf{5} & 0.505 & 0.416 & 0.616 & 0.614 \\ 
			\textbf{6} & 0.518 & 0.534 & 0.638 & 0.729 \\ 
			\textbf{7} & 0.385 & 0.320 & 0.423 & 0.407 \\ \midrule
			Correlation(Avg $\pm$ std) &   0.51$\pm$0.09 & 0.51$\pm$0.13 & 0.60$\pm$0.12 & 0.63$\pm$0.12     \\ \bottomrule
		\end{tabularx}
	\end{table*}
	
	\begin{table*}[!htbp]  
		\caption{Correlation values between the generated and recorded EEG responses of 2-4 repetitions of 3-bit wide pulse, for all 7 subjects (PR study).}
		\label{prW3Table}
		\begin{tabularx}{\textwidth} {@{}l*{2}{C}c@{}}
			\toprule
			\textbf{Subject No}  & \textbf{2} & \textbf{3} & \textbf{4}     \\ \midrule
			\textbf{1} & 0.603 & 0.557 & 0.656 \\ 
			\textbf{2} & 0.725 & 0.822 & 0.804 \\ 
			\textbf{3} & 0.596 & 0.576 & 0.484 \\ 
			\textbf{4} & 0.406 & 0.696 & 0.724 \\ 
			\textbf{5} & 0.519 & 0.634 & 0.422 \\ 
			\textbf{6} & 0.525 & 0.712 & 0.688 \\ 
			\textbf{7} & 0.688 & 0.722 & 0.820 \\ \midrule
			Correlation(Avg $\pm$ std) &   0.58$\pm$0.11 & 0.67$\pm$0.09 & 0.66$\pm$0.15\\ \bottomrule
		\end{tabularx}
	\end{table*}

	
	\begin{table*}[!htbp]
		\caption{ Repeated measures ANOVA results on the correlations obtained for 1-9 bit wide pulses. }
		\centering
		\label{pwStudyANOVA}
		\begin{tabular}{|c|c|c|c|c|c|}
			\toprule
			&  \textbf{SumSq} &  \textbf{DF} & \textbf{MeanSq} &  \textbf{F} &  \textbf{pValue} \\ \midrule
			\textbf{(Intercept):}  \textbf{Measurements} & 0.7428 &  8 & 0.0928 & 7.6067 & 1.6429e-06   \\ 
			\textbf{Error} \textbf{ (Measurements) }     & 0.5859 & 48 & 0.0122 &        &              \\ \bottomrule
		\end{tabular} 
	\end{table*}
	
	\begin{table*}[!htbp]
		\caption{ Repeated measures ANOVA results on the correlations obtained for 1-9 bit separations between two 1-bit wide pulses. }
		\label{ps1ANOVA}
		\centering	
		\begin{tabular}{|c|c|c|c|c|c|}
			\toprule
			& { \textbf{SumSq}} & {\textbf{DF}} & {\textbf{MeanSq}} & {\textbf{F}} & {\textbf{pValue}} \\ \midrule
			\textbf{(Intercept):Measurements} & 0.4488 &  8 & 0.0561 & 2.7829 & 0.01296     \\ \hline
			\textbf{Error(Measurements) }     & 0.9676 & 48 & 0.0202 &        &              \\ \bottomrule
		\end{tabular} 
	\end{table*}
	
	\begin{table*}[!htbp]
		\caption{ Repeated measures ANOVA results on the correlations obtained for 1-9 bit separations between two 2-bit wide pulses. }
		\label{ps2ANOVA} 
		\centering	
		\begin{tabular}{|c|c|c|c|c|c|}
			\toprule
			& {\textbf{SumSq}} & {\textbf{DF}} & {\textbf{MeanSq}} & {\textbf{F}} & { \textbf{pValue}} \\ \midrule
			\textbf{(Intercept):Measurements} & 0.77103 &  8 & 0.09637 & 7.3629 & 2.4356e-06      \\ 
			\textbf{Error(Measurements) }     & 0.62831 & 48 & 0.01309 &        &                 \\ \bottomrule
		\end{tabular} 
	\end{table*}

	\begin{table*}[!htbp]
		\caption{ Repeated measures ANOVA results on the correlations obtained for 2-6 repetitions of 1-bit wide pulse. }
		\label{prW1ANOVA}
		\centering	
		\begin{tabular}{|c|c|c|c|c|c|}
			\toprule
			& {\textbf{SumSq}} & {\textbf{DF}} & {\textbf{MeanSq}} & {\textbf{F}} & {\textbf{pValue}} \\ \midrule
			\textbf{(Intercept):Measurements} & 0.32518 &  4 & 0.081296 & 3.2185 & 0.029995      \\ \hline
			\textbf{Error(Measurements) }     & 0.60622 & 24 & 0.025259 &        &                 \\ \bottomrule
		\end{tabular} 
	\end{table*}

	\begin{table*}[!htbp]
		\caption{ Repeated measures ANOVA results on the correlations obtained for 2-5 repetitions of 2-bit wide pulse. }
		\label{prW2ANOVA}
		\centering	
		\begin{tabular}{|c|c|c|c|c|c|}
			\toprule
			& {\textbf{SumSq}} & {\textbf{DF}} & { \textbf{MeanSq}} & {\textbf{F}} & {\textbf{pValue}} \\ \midrule
			\textbf{(Intercept):Measurements} & 0.080398 &  3 & 0.026799 & 6.0311 & 0.0049883      \\ \hline
			\textbf{Error(Measurements) }     & 0.079984 & 18 & 0.0044436 &        &                 \\ \bottomrule
		\end{tabular} 
	\end{table*}

	\begin{table*}[!htbp]
		\caption{ Repeated measures ANOVA results on the correlations obtained for 2-4 repetitions of 3-bit wide pulse. }
		\centering
		\label{prW3ANOVA}
		\begin{tabular}{|c|c|c|c|c|c|}
			\toprule
			& { \textbf{SumSq}} & { \textbf{DF}} & { \textbf{MeanSq}} & { \textbf{F}} & {\ \textbf{pValue}} \\ \midrule
			\textbf{(Intercept):Measurements} & 0.034933 &  2 & 0.017466 & 2.1593 & 0.15812      \\ \hline
			\textbf{Error(Measurements) }     & 0.097065 & 12 & 0.008088 &        &              \\ \bottomrule
		\end{tabular} 
	\end{table*}


	\begin{table*}[!htbp]
		\caption{ Results of the pairwise paired t-tests on the correlations obtained for 1-9 bit wide pulses.  }
		\label{pwPairWiseResult}
		\centering
		\begin{tabular}{|c|c|c|c|c|c|c|c|c|c|}
			\hline
			{\textbf{PW}} 
			& {\textbf{1}} & { \textbf{2}} & { \textbf{3}} & { \textbf{4}} & { \textbf{5}} & {\textbf{6}} & {\textbf{7}} & {\textbf{8}} & {\textbf{9}} \\ \hline
			{\textbf{1}} &   & 0 & 1 & 1 & 1 & 1 & 1 & 1 & 0 \\ \hline
			{\textbf{2}} & 0 &   & 0 & 0 & 1 & 1 & 1 & 1 & 0 \\ \hline
			{\textbf{3}} & 1 & 0 &   & 0 & 1 & 1 & 1 & 0 & 0 \\ \hline
			{\textbf{4}} & 1 & 0 & 0 &   & 1 & 0 & 0 & 0 & 0 \\ \hline
			{\textbf{5}} & 1 & 1 & 1 & 1 &   & 0 & 1 & 0 & 0 \\ \hline
			{\textbf{6}} & 1 & 1 & 1 & 0 & 0 &   & 0 & 0 & 1 \\ \hline
			{\textbf{7}} & 1 & 1 & 1 & 0 & 1 & 0 &   & 0 & 0 \\ \hline
			{\textbf{8}} & 1 & 1 & 0 & 0 & 0 & 0 & 0 &   & 0 \\ \hline
			{\textbf{9}} & 0 & 0 & 0 & 0 & 0 & 1 & 0 & 0 &   \\ \hline
			
		\end{tabular}
	\end{table*}

	\begin{table*}[!htbp]
		\caption{ Results of the pairwise paired t-tests on the correlations obtained for 1-9 bit separations between two 1-bit wide pulses. }
		\label{ps1PairWisePResult}
		\centering
		\begin{tabular}{|c|c|c|c|c|c|c|c|c|c|}
			\hline
			{ \textbf{PS1}} 
			& {\textbf{1}} & { \textbf{2}} & {\textbf{3}} & {\textbf{4}} & { \textbf{5}} & {\textbf{6}} & { \textbf{7}} & {\textbf{8}} & { \textbf{9}} \\ \hline
			{\textbf{1}} & 0  &   0  &   0  &   0  &   0  &   0  &   0  &   0  &   0 \\ \hline
			{\textbf{2}} & 0  &      &   0  &   0  &   0  &   1  &   0  &   1  &   0 \\ \hline
			{\textbf{3}} & 0  &   0  &      &   1  &   1  &   1  &   0  &   1  &   1 \\ \hline
			{\textbf{4}} & 0  &   0  &   1  &      &   0  &   0  &   0  &   0  &   0 \\ \hline
			{\textbf{5}} & 0  &   0  &   1  &   0  &      &   0  &   0  &   0  &   0 \\ \hline
			{\textbf{6}} & 0  &   1  &   1  &   0  &   0  &      &   0  &   0  &   0 \\ \hline
			{\textbf{7}} & 0  &   0  &   0  &   0  &   0  &   0  &      &   0  &   0 \\ \hline
			{\textbf{8}} & 0  &   1  &   1  &   0  &   0  &   0  &   0  &      &   0 \\ \hline
			{\textbf{9}} & 0  &   0  &   1  &   0  &   0  &   0  &   0  &   0  &     \\ \hline
		\end{tabular}
	\end{table*}

	\begin{table*}[!htbp]
		\caption{ Results of the pairwise paired t-tests on the correlations obtained for 1-9 bit separations between two 2-bit wide pulses.}
		\centering
		\label{ps2PairWisePResult} 
		\begin{tabular}{|c|c|c|c|c|c|c|c|c|c|}
			\hline
			{\textbf{PS2}} 
			& {\textbf{1}} & {\textbf{2}} & {\textbf{3}} & {\textbf{4}} & {\textbf{5}} & {\textbf{6}} & {\textbf{7}} & {\textbf{8}} & {\textbf{9}} \\ \hline
			{\textbf{1}} &    &   1  &   1  &   1  &   1  &   1  &   0  &   1  &   1 \\ \hline
			{\textbf{2}} & 1  &      &   1  &   1  &   0  &   0  &   0  &   0  &   1 \\ \hline
			{\textbf{3}} & 1  &   1  &      &   0  &   0  &   0  &   0  &   0  &   0 \\ \hline
			{\textbf{4}} & 1  &   1  &   0  &      &   0  &   0  &   0  &   0  &   0 \\ \hline
			{\textbf{5}} & 1  &   0  &   0  &   0  &      &   0  &   0  &   0  &   0 \\ \hline
			{\textbf{6}} & 1  &   0  &   0  &   0  &   0  &      &   0  &   0  &   0 \\ \hline
			{\textbf{7}} & 0  &   0  &   0  &   0  &   0  &   0  &      &   0  &   0 \\ \hline
			{\textbf{8}} & 1  &   0  &   0  &   0  &   0  &   0  &   0  &      &   0 \\ \hline
			{\textbf{9}} & 1  &   1  &   0  &   0  &   0  &   0  &   0  &   0  &     \\ \hline
		\end{tabular}
	\end{table*}
	
	\begin{table*}
		\caption{ Results of the pairwise paired t-tests on the correlations obtained for 2-6 repetitions of 1-bit wide pulse. }
		\label{prW1tTest}
		\centering
		\begin{tabular}{|c|c|c|c|c|c|}
			\hline
			{\textbf{PR1}} 
			& {\textbf{2}} & {\textbf{3}} & {\textbf{4}} & {\textbf{5}}  & {\textbf{6}}\\ \hline
			{\textbf{2}} &    &   0  &   1  &   0  &   0 \\ \hline
			{\textbf{3}} & 0  &      &   0  &   0  &   0 \\ \hline
			{\textbf{4}} & 1  &   0  &      &   0  &   0 \\ \hline
			{\textbf{5}} & 0  &   0  &   0  &      &   0 \\ \hline
			{\textbf{6}} & 0  &   0  &   0  &   0  &     \\ \hline
		\end{tabular}
	\end{table*}
	
	\begin{table*}[!htbp]
		\caption{ Results of the pairwise paired t-tests on the correlations obtained for 2-5 repetitions of 2-bit wide pulse. }
		\label{prW2tTest}
		\centering
		\begin{tabular}{|c|c|c|c|c|}
			\hline
			{\textbf{PR2}} 
			& {\textbf{2}} & {\textbf{3}} & {\textbf{4}} & {\textbf{5}}  \\ \hline
			{\textbf{2}} &    &   0  &   0  &   1  \\ \hline
			{\textbf{3}} & 0  &      &   1  &   1  \\ \hline
			{\textbf{4}} & 0  &   1  &      &   0  \\ \hline
			{\textbf{5}} & 1  &   1  &   0  &      \\ \hline
		\end{tabular}
	\end{table*}
	
	\begin{table*}
		\caption{Results of the pairwise paired t-test on the correlations obtained for 2-4 repetitions of 3-bit wide pulse. }
		\label{prW3tTest}
		\centering
		\begin{tabular}{|c|c|c|c|c|}
			\hline
			{\textbf{PR3}} & {\textbf{2}} & {\textbf{3}} & {\textbf{4}}  \\ \hline
			{\textbf{2}} &    &   0  &   0 \\ \hline
			{\textbf{3}} & 0  &      &   0 \\ \hline
			{\textbf{4}} & 0  &   0  &     \\ \hline
		\end{tabular}
	\end{table*}

	\begin{table*}
		\caption{ P-Values of the pairwise paired t-tests on the correlations obtained for 1-9 bit wide pulses. }
		\label{pwPairWisePval}
		\centering
		\begin{tabular}{|c|c|c|c|c|c|c|c|c|c|}
			\hline
			{\textbf{PW}} 
			& {\textbf{1}} & { \textbf{2}} & {\textbf{3}} & {\textbf{4}} & {\textbf{5}} & {\textbf{6}} & { \textbf{7}} & {\textbf{8}} & { \textbf{9}} \\ \hline
			{ \textbf{1}} &         &  0.7479 &  0.0016 & 0.0160 & 0.0006 & 0.0096 & 0.0009 & 0.0119 & 0.0658 \\ \hline
			{ \textbf{2}} & 0.7479  &         &  0.0721 & 0.0696 & 0.0002 & 0.0034 & 0.0030 & 0.0067 & 0.1059 \\ \hline
			{ \textbf{3}} & 0.0016  &  0.0721 &         & 0.3963 & 0.0075 & 0.0350 & 0.0118 & 0.0764 & 0.6193 \\ \hline
			{ \textbf{4}} & 0.0160  &  0.0696 &  0.3963 &        & 0.0245 & 0.2888 & 0.5049 & 0.6484 & 0.6943 \\ \hline
			{ \textbf{5}} & 0.0006  &  0.0002 &  0.0075 & 0.0245 &        & 0.4510 & 0.0309 & 0.1023 & 0.0621 \\ \hline
			{ \textbf{6}} & 0.0096  &  0.0034 &  0.0350 & 0.2888 & 0.4510 &        & 0.4359 & 0.1616 & 0.0036 \\ \hline
			{ \textbf{7}} & 0.0009  &  0.0030 &  0.0118 & 0.5049 & 0.0309 & 0.4359 &        & 0.9320 & 0.2519 \\ \hline
			{ \textbf{8}} & 0.0119  &  0.0067 &  0.0764 & 0.6484 & 0.1023 & 0.1616 & 0.9320 &        & 0.1536 \\ \hline
			{ \textbf{9}} & 0.0658  &  0.1059 &  0.6193 & 0.6943 & 0.0621 & 0.0036 & 0.2519 & 0.1536 &        \\ \hline
		\end{tabular}
	\end{table*}
	
	\begin{table*}
		\caption{ P-Values of the pairwise paired t-tests on the correlations obtained for 1-9 bit separations between two 1-bit wide pulses. }
		\label{ps1PairWisePval}
		\centering
		\begin{tabular}{|c|c|c|c|c|c|c|c|c|c|}
			\hline
			{\textbf{PS1}} 
			& {\textbf{1}} & {\textbf{2}} & { \textbf{3}} & { \textbf{4}} & {\textbf{5}} & {\textbf{6}} & { \textbf{7}} & {\textbf{8}} & { \textbf{9}} \\ \hline
			{\textbf{1}} &         &  0.8899  &  0.8848  &  0.3044  &  0.1289  &  0.0552  &  0.4724  &  0.0866 &   0.1423 \\ \hline
			{\textbf{2}} & 0.8899  &          &  0.7097  &  0.1669  &  0.1402  &  0.0270  &  0.3277  &  0.0250 &   0.0824 \\ \hline
			{\textbf{3}} & 0.8848  &  0.7097  &          &  0.0236  &  0.0131  &  0.0079  &  0.1025  &  0.0036 &   0.0265 \\ \hline
			{\textbf{4}} & 0.3044  &  0.1669  &  0.0236  &          &  0.5816  &  0.3783  &  0.5275  &  0.2726 &   0.4447 \\ \hline
			{\textbf{5}} & 0.1289  &  0.1402  &  0.0131  &  0.5816  &          &  0.6125  &  0.3906  &  0.9343 &   0.8815 \\ \hline
			{\textbf{6}} & 0.0552  &  0.0270  &  0.0079  &  0.3783  &  0.6125  &          &  0.1633  &  0.6196 &   0.7923 \\ \hline
			{\textbf{7}} & 0.4724  &  0.3277  &  0.1025  &  0.5275  &  0.3906  &  0.1633  &          &  0.2023 &   0.0596 \\ \hline
			{\textbf{8}} & 0.0866  &  0.0250  &  0.0036  &  0.2726  &  0.9343  &  0.6196  &  0.2023  &         &   0.9131 \\ \hline
			{\textbf{9}} & 0.1423  &  0.0824  &  0.0265  &  0.4447  &  0.8815  &  0.7923  &  0.0596  &  0.9131 &          \\ \hline
			
		\end{tabular}
	\end{table*}

	\begin{table*}
		\caption{P-Values of the pairwise paired t-tests on the correlations obtained for 1-9 bit separations between two 2-bit wide pulses.  }
		\label{ps2PairWisePval}
		\centering
		
		\begin{tabular}{|c|c|c|c|c|c|c|c|c|c|}
			\hline
			{\textbf{PS2}} 
			& {\textbf{1}} & {\textbf{2}} & {\textbf{3}} & {\textbf{4}} & {\textbf{5}} & {\textbf{6}} & {\textbf{7}} & {\textbf{8}} & {\textbf{9}} \\ \hline
			{\textbf{1}} &         &  0.0011  &  0.0008  &  0.0008  &  0.0041  &  0.0056  &  0.0502  &  0.0064  &  0.0025 \\ \hline
			{\textbf{2}} & 0.0011  &          &  0.0249  &  0.0197  &  0.1431  &  0.1176  &  0.7097  &  0.0880  &  0.0448 \\ \hline
			{\textbf{3}} & 0.0008  &  0.0249  &          &  0.6112  &  0.6593  &  0.8567  &  0.2089  &  0.9616  &  0.5160 \\ \hline
			{\textbf{4}} & 0.0008  &  0.0197  &  0.6112  &          &  0.2391  &  0.5147  &  0.0980  &  0.6616  &  0.5849 \\ \hline
			{\textbf{5}} & 0.0041  &  0.1431  &  0.6593  &  0.2391  &          &  0.5676  &  0.1661  &  0.6437  &  0.1563 \\ \hline
			{\textbf{6}} & 0.0056  &  0.1176  &  0.8567  &  0.5147  &  0.5676  &          &  0.1381  &  0.8580  &  0.4435 \\ \hline
			{\textbf{7}} & 0.0502  &  0.7097  &  0.2089  &  0.0980  &  0.1661  &  0.1381  &          &  0.1882  &  0.0527 \\ \hline
			{\textbf{8}} & 0.0064  &  0.0880  &  0.9616  &  0.6616  &  0.6437  &  0.8580  &  0.1882  &          &  0.3448 \\ \hline
			{\textbf{9}} & 0.0025  &  0.0448  &  0.5160  &  0.5849  &  0.1563  &  0.4435  &  0.0527  &  0.3448  &         \\ \hline
			
		\end{tabular}
	\end{table*}

	\begin{table*}
		\caption{ P-Values of the pairwise paired t-tests on the correlations obtained for 2-6 repetitions of 1-bit wide pulse. }
		\label{prW1Pvalues}
		\centering
		\begin{tabular}{|c|c|c|c|c|c|}
			\hline
			{\textbf{PR1}} 
			& {\textbf{2}} & {\textbf{3}} & { \textbf{4}} & { \textbf{5}} & { \textbf{6}}   \\ \hline
			{\textbf{2}} &         &  0.2374  &  0.0481  &  0.0754  &  0.0817 \\ \hline
			{\textbf{3}} & 0.2374  &          &  0.1591  &  0.1032  &  0.2768 \\ \hline
			{\textbf{4}} & 0.0481  &  0.1591  &          &  0.6400  &  0.7765 \\ \hline
			{\textbf{5}} & 0.0754  &  0.1032  &  0.6400  &          &  0.8157 \\ \hline
			{\textbf{6}} & 0.0817  &  0.2768  &  0.7765  &  0.8157  &         \\ \hline
		\end{tabular}
	\end{table*}

	\begin{table*}
		\caption{ P-Values of the pairwise paired t-tests on the correlations obtained for 2-5 repetitions of 2-bit wide pulse. }
		\label{prW2pVal}
		\centering
		\begin{tabular}{|c|c|c|c|c|c|}
			\hline
			{\textbf{PR2}} & {\textbf{2}} & {\textbf{3}} & {\textbf{4}} & {\textbf{5}}  \\ \hline
			{\textbf{2}} &         &  0.9274  &  0.0837  &  0.0174  \\ \hline
			{\textbf{3}} & 0.9274  &          &  0.0144  &  0.0061  \\ \hline
			{\textbf{4}} & 0.0837  &  0.0144  &          &  0.1916  \\ \hline
			{\textbf{5}} & 0.0174  &  0.0061  &  0.1916  &          \\ \hline
			
		\end{tabular}
	\end{table*}
	
	\begin{table*}
		\caption{P-Values of the pairwise paired t-tests on the correlations obtained for 2-4 repetitions of 3-bit wide pulse. }
		\label{prW3pVal}
		\centering
		\begin{tabular}{|c|c|c|c|c|c|}
			\hline
			{\textbf{PR3}} & {\textbf{2}} & {\textbf{3}} & {\textbf{4}}  \\ \hline
			{\textbf{2}} &         &  0.0802  &  0.2257 \\ \hline
			{\textbf{3}} & 0.0802  &          &  0.6917 \\ \hline
			{\textbf{4}} & 0.2257  &  0.6917  &         \\ \hline
		\end{tabular}
	\end{table*}

\end{document}